\documentclass[
reprint,
superscriptaddress,
showpacs,
preprintnumbers,
nofootinbib,
nobibnotes,
amsmath,
amssymb, 
aps,
prd,
floatfix
]{revtex4-2}

\pdfoutput=1 

\usepackage[utf8]{inputenc}
\usepackage[normalem]{ulem}
\usepackage{graphicx}
\usepackage{dcolumn}
\usepackage{bm}
\usepackage{color}
\usepackage{xcolor}
\usepackage{url}
\usepackage{enumerate}

\usepackage[newfloat,frozencache,cachedir=.]{minted}
\usepackage{listings}
\usepackage{tikz}
\usepackage{tcolorbox}
\tcbuselibrary{skins}
\tcbuselibrary{listings}
\tcbuselibrary{breakable}

\usepackage{slashed}
\usepackage{multirow}
\usepackage{mathrsfs} 
\usepackage{amsmath}
\usepackage{xspace}

\usepackage{bbold}
\usepackage{mathrsfs}
\usepackage{braket}
\usepackage{physics}
\usepackage{booktabs}
\usepackage{longtable}
\usepackage{multirow}
\usepackage{enumitem}   

\usepackage{fontawesome5} 

\usepackage[colorlinks=true,allcolors=purple]{hyperref}
\usepackage{cleveref}

\definecolor{orcidlogocol}{HTML}{A6CE39}

\definecolor{monokaibg}{HTML}{272822}
\definecolor{monokaifg}{rgb}{0.97,0.97,0.95}
\definecolor{mygrzay}{gray}{0.5}
\definecolor{azure}{rgb}{0.97, 1.0, 1.0}

\definecolor{blue-violet}{rgb}{0.33, 0.17, 0.89}

\newminted{python}{
linenos,
frame=single,
stepnumber=1,
numbersep=0pt,
breaklines,
tabsize=4,
autogobble,
framesep=1mm,
baselinestretch=1.2,
bgcolor=azure,
fontsize=\footnotesize}
\setmintedinline[python]{bgcolor=azure,fontsize=\small}

\newminted{shell-session}{
linenos,
frame=single,
stepnumber=1,
numbersep=0pt,
breaklines,
tabsize=4,
autogobble,
framesep=1mm,
baselinestretch=1.2,
bgcolor=monokaibg,
fontsize=\footnotesize}
\setmintedinline[shell-session]{bgcolor=monokaibg,fontsize=\small}

\SetupFloatingEnvironment{listing}{name=Code}

\newtcblisting{inputfile}[1][]{enhanced,listing only,
boxrule=0.4pt,left=3mm,right=2mm,top=1mm,bottom=1mm,
colback=white,
colframe=black,
boxed title style={size=small,colback=white!82!black,boxrule=0.4pt},
fonttitle=\ttfamily\footnotesize,coltitle=black,
sharp corners,rounded corners=southeast,arc is angular,arc=3mm,
underlay={%
  \path[fill=white!70!black] ([yshift=3mm]interior.south east)--++(-0.4,-0.1)--++(0.1,-0.2);
  \path[draw=black,shorten <=-0.05mm,shorten >=-0.05mm] ([yshift=3mm]interior.south east)--++(-0.4,-0.1)--++(0.1,-0.2);
  },
drop fuzzy shadow,attach boxed title to top right={yshift=-2mm, xshift=-4mm},#1}

\makeatletter 
    
\renewcommand\onecolumngrid{
\do@columngrid{one}{\@ne}%
\def\set@footnotewidth{\onecolumngrid}
\def\footnoterule{\kern-6pt\hrule width 1.5in\kern6pt}%
}

\renewcommand\twocolumngrid{
        \def\footnoterule{
        \dimen@\skip\footins\divide\dimen@\thr@@
        \kern-\dimen@\hrule width.5in\kern\dimen@}
        \do@columngrid{mlt}{\tw@}
}%

\makeatother  

\newcommand{\refref}[1]{Ref.~\cite{#1}}

\renewcommand{\phi}{\varphi}

\newcounter{CommentCount}
\definecolor{MH}{rgb}{0.0,0.6,9}
\definecolor{palatinate}{rgb}{0.494, 0.192, 0.482}

\definecolor{teal}{HTML}{008080}

\newcommand{\dngen}{DarkNews-generator\xspace}
\newcommand{\darknus}{\texttt{DarkNews}\xspace}
\newcommand{\vegas}{\texttt{vegas}\xspace}
\newcommand{\pandas}{\texttt{pandas}\xspace}
\newcommand{\numpy}{\texttt{numpy}\xspace}
\newcommand{\genie}{\texttt{Genie}\xspace}
\newcommand{\achilles}{\texttt{Achilles}\xspace}
\newcommand{\leptoninjector}{\texttt{LeptonInjector}\xspace}
\newcommand{\python}{\texttt{Python}\xspace}

\newcommand{\orcid}[1]{\href{https://orcid.org/#1}{(\textcolor[HTML]{A6CE39}{\faOrcid} #1)}}

\begin{document}

\title{\darknus: a \python-based event generator for heavy neutral lepton production in neutrino-nucleus scattering}

\author{Asli M. Abdullahi}
\email{asli@fnal.gov}
\thanks{\orcid{0000-0002-6122-4986}}
\affiliation{Theoretical Physics Department, Fermi National Accelerator Laboratory, Batavia, IL 60510, USA}

\author{Jaime Hoefken Zink}
\email{jaime.hoefkenzink2@unibo.it}
\thanks{\orcid{0000-0002-4086-2030}}
\affiliation{Dipartimento di Fisica e Astronomia, Universit\`a di Bologna, via Irnerio 46, 40126 Bologna, Italy}
\affiliation{INFN, Sezione di Bologna, viale Berti Pichat 6/2, 40127 Bologna, Italy}

\author{Matheus Hostert}
\email{mhostert@perimeterinstitute.ca}
\thanks{\orcid{0000-0002-9584-8877}}
\affiliation{Perimeter Institute for Theoretical Physics, Waterloo, ON N2J 2W9, Canada}
\affiliation{School of Physics and Astronomy, University of Minnesota, Minneapolis, MN 55455, USA}
\affiliation{William I. Fine Theoretical Physics Institute, School of Physics and Astronomy, University of Minnesota, Minneapolis, MN 55455, USA}

\author{Daniele Massaro}
\email{daniele.massaro5@unibo.it}
\thanks{\orcid{0000-0002-1013-3953}}
\affiliation{Dipartimento di Fisica e Astronomia, Universit\`a di Bologna, via Irnerio 46, 40126 Bologna, Italy}
\affiliation{INFN, Sezione di Bologna, viale Berti Pichat 6/2, 40127 Bologna, Italy}
\affiliation{Centre for Cosmology, Particle Physics and Phenomenology (CP3), Universit\'e Catholique de Louvain, B-1348 Louvain-la-Neuve, Belgium}

\author{Silvia Pascoli}
\email{silvia.pascoli@unibo.it}
\thanks{\orcid{0000-0002-2958-456X}}
\affiliation{Dipartimento di Fisica e Astronomia, Universit\`a di Bologna, via Irnerio 46, 40126 Bologna, Italy}
\affiliation{INFN, Sezione di Bologna, viale Berti Pichat 6/2, 40127 Bologna, Italy}
\affiliation{CERN, Theoretical Physics Department, Geneva, Switzerland}

\date{\today}

\begin{abstract}
We introduce \darknus, a lightweight \python-based Monte-Carlo generator for beyond-the-Standard-Model neutrino-nucleus scattering. 
The generator handles the production and decay of heavy neutral leptons via additional vector or scalar mediators, as well as through transition magnetic moments. 
\darknus samples pre-computed neutrino-nucleus upscattering cross sections and heavy neutrino decay rates to produce dilepton and single-photon events in accelerator neutrino experiments.
We present two case studies with differential distributions for models that can explain the MiniBooNE excess.
The aim of this code is to aid the neutrino theory and experimental communities in performing searches and sensitivity studies for new particles produced in neutrino upscattering.
\end{abstract}

\maketitle

\section{Introduction}

The discovery of physics beyond the Standard Model (BSM) is among the highest priority goals of modern experimental particle physics.
While searches for heavy new physics continue at the Large Hadron Collider, other low-energy experiments continue to make progress in searching for lighter, but weakly-coupled particles~\cite{Alexander:2016aln,Agrawal:2021dbo}.
Among these, neutrino experiments stand out for their high intensity and detector capabilities.
The sheer number of protons delivered on target produces an intense neutrino beam, and possibly large numbers of hypothetical particles.
Together with the fact that neutrinos interact very little with matter, accelerator neutrino experiments are an ideal environment to look for the feeble interactions of hidden sector particles.

In this context, light dark sectors are particularly relevant as they may contain particles that propagate freely between the proton target and the neutrino detector before depositing some amount of energy either through decays or interactions with matter. 
Numerous studies in the literature have explored these features to set limits on light dark matter particles, light scalar and vector bosons, and heavy neutral leptons~\cite{Batell:2009di,Essig:2010gu,deNiverville:2011it,Gninenko:2011uv,deNiverville:2012ij} (for recent overviews, see Refs.~\cite{Buonocore:2019esg,Arguelles:2019xgp,Abdullahi:2022jlv}).
Several neutrino detectors have come online in the past decade, and several more will become operational in the near future.
These include the near and far detectors of the short-baseline program at Fermilab~\cite{MicroBooNE:2015bmn}, the upgraded near detector of T2K~\cite{T2K:2019bbb}, and, on a more distant time scale, Hyper-Kamiokande~\cite{Hyper-Kamiokande:2016dsw} and DUNE~\cite{DUNE:2020ypp}.
Several new proposals are also currently under study~\cite{Toups:2022knq}. 
Exploring their capabilities to probe light new physics will require efforts from both the experimental and theoretical particle physics communities. On the theory side, the provision of theoretical targets and tools to simulate the effects of light new physics at experiments will be crucial.

New interactions shared between neutrinos and dark sector particles are usually also shared by other charged particles in the SM.
One notable exception is given by the neutrino portal
\begin{equation}
    \mathcal{L} \supset \overline{L} \widetilde{H} N + \mathcal{L}_{N{\rm -int}},
\end{equation}
where new interactions felt by a heavy neutrino $N$ can be shared with the Standard Model neutrinos $\nu$ via mixing.
In the absence of new forces, $N$ interacts with SM particles via the Weak interactions, further suppressed by mixing elements $U_{\alpha N}$ which are experimentally constrained to be very small. However, even very feeble interactions between $N$ and other dark sector states can dominate over the "weaker-than-Weak'' couplings between $N$ and the SM W and Z bosons.
In this case, it is possible that BSM interactions of $N$ can dominate its decay modes and lead to new production mechanisms.

One new production mechanism that has gained interest recently is the production of heavy neutrinos by neutrinos scattering off of ordinary matter. 
Their subsequent decays to charged particles via the new force have been used in a number of phenomenological papers.
This new force may be mediated by, for example, a dark photon or dark scalar, or even by the SM photon itself if transition magnetic moments between light and heavy neutrinos exist.
These models have a relatively simple phenomenology in neutrino experiments: neutrino scattering with a nuclear target and potentially displaced decay signatures with photons, charged lepton pairs, or meson final states.

Currently, to model new physics in neutrino-nucleus scattering the community often relies on custom-made generators, or the manual addition of new interactions and particles to existing neutrino generators e.g. \genie~\cite{Andreopoulos:2009rq,GENIE:2021zuu}, \texttt{NuWro}~\cite{Golan:2012wx}, \texttt{Neut}~\cite{Hayato:2021heg}, and \texttt{GiBUU}~\cite{Buss:2011mx}.
This is to be contrasted with collider physics, where decades-long efforts have produced complete and automated Monte-Carlo tools that go from a user-defined Lagrangian input, e.g. FeynRules~\cite{Christensen:2008py,Alloul:2013bka}, to matrix elements in the standard UFO file format~\cite{Degrande:2011ua} with a universal interface with event generators (see \refref{Campbell:2022qmc} for a recent overview of Monte-Carlo generators).
Similarly, in the intensity frontier, exploration of dark matter models has followed a similar route with programs like \texttt{MicrOMEGAs}~\cite{Belanger:2006is,Belanger:2013oya,Belanger:2014vza,Belanger:2018ccd}, \texttt{MadDM}~\cite{Backovic:2013dpa,Backovic:2015cra,Ambrogi:2018jqj,Arina:2021gfn}, as well as the more recent \texttt{MadDump}~\cite{Buonocore:2018xjk} for dark sectors in beam dump-style experiments.

In neutrino physics, these efforts are rather recent.
Tools like \leptoninjector~\cite{IceCube:2020tcq} specialize in providing a detailed description of neutrino interactions in arbitrary and complex detector geometries at neutrino telescopes as well as at accelerator neutrino experiments.
Implementing new physics models in \leptoninjector is possible (as done recently in Ref~\cite{Kamp:2022bpt}), but also relies on a case-by-case approach.
On the other hand, the recently-developed tool \achilles~\cite{Isaacson:2021xty,Isaacson:2022cwh} aims to combine a state-of-the-art description of nuclear physics in lepton-nucleus interactions with automated particle physics tools. 
These tools have recently gained great interest from both the theoretical and experimental community for their interface with tools like FeynRules. 
In view of upcoming large-scale projects like DUNE and Hyper-Kamiokande, they will be of great relevance for the phenomenology community studying new physics models at GeV neutrino experiments.
Nevertheless, there is a need for simpler tools that can provide ready-to-use simulations of specific new physics models.
This is especially true when it comes to probing new-physics solutions to the short-baseline puzzle.
Such generators are also of great importance for phenomenologists studying the sensitivity of upcoming experiments to models of new physics and for whom the use of more sophisticated generators is time-consuming. 
In this article, we present a simple tool that aims to do just this for a range of heavy neutral lepton (HNL) models. 

This article is divided as follows: in \Cref{sec:generator}, we introduce the generator and describe its pipeline, inputs, and outputs.
In \Cref{sec:models}, we discuss the new physics models already implemented in the code, providing a few example simulations with DarkNews in \Cref{sec:examples}.
Finally, we conclude with \Cref{sec:conclusions}. Tables of input and output formats are given in \Cref{sec:appendix}.

\section{The Generator}\label{sec:generator}
\begin{figure}[t]
    {\centering \href{https://github.com/mhostert/DarkNews-generator}{\includegraphics[width=0.22\textwidth]{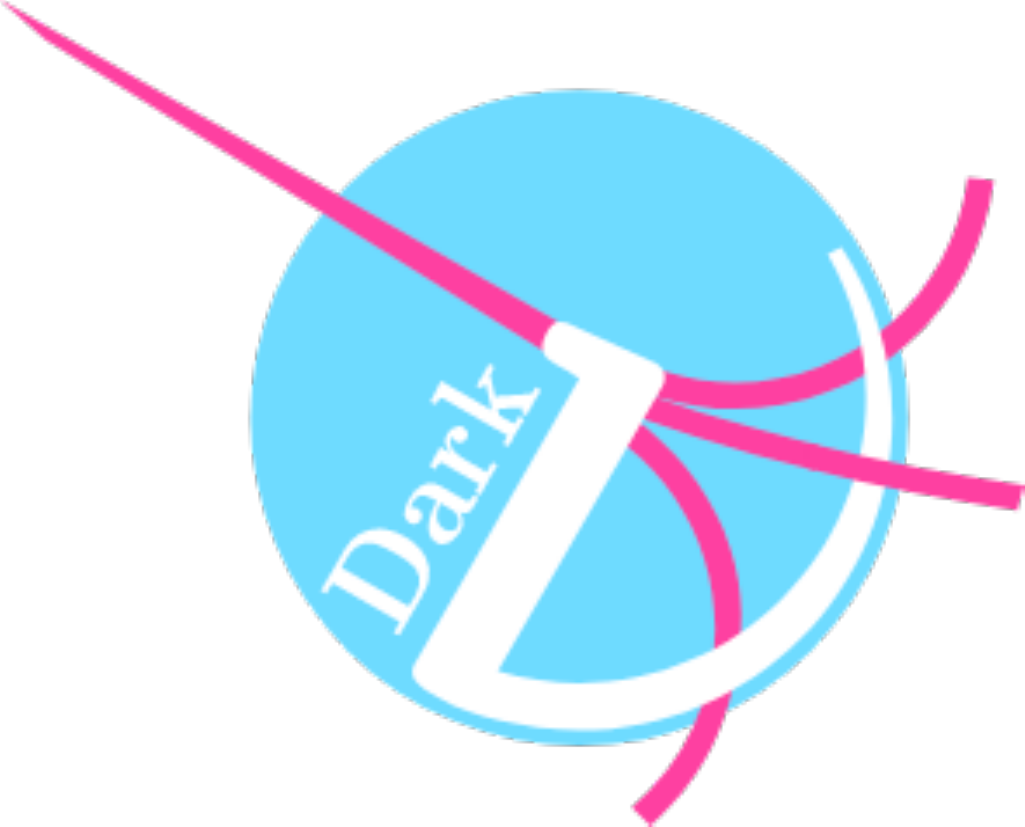}}\\
    \begin{minipage}{0.4\textwidth}
    \vspace{2ex}
    \begin{minted}{shell-session}
            pip install DarkNews
    \end{minted}
    \end{minipage}}
    \caption{The \darknus logo and the installation command in systems where pip is linked to the desired version of \python. \label{fig:darknus_install}}
\end{figure}

\darknus provides a simple, fast, and \python-based generator that simulates neutrino-induced production and decay of heavy neutrinos inside neutrino detectors.
It builds on \vegas~\cite{Lepage:1977sw,Lepage:2020tgj} to sample from pre-computed, analytical differential cross sections and decay rates, using standard algorithms for adaptive and stratified sampling.
The main scope of the generator is in the modeling of coherent neutrino-nucleus scattering followed by the decay of HNLs endowed with additional interactions.
We also provide cross sections for incoherent scattering on nucleons, but we do not model any nuclear physics.
The models that are currently implemented are discussed in \Cref{sec:models}.
Implementation of other new physics scenarios in \dngen can be pursued on a case-by-case basis.

The code takes as input: neutrino fluxes, model parameters, and nuclear targets, and provides as an output a set of weighed events with all particles involved, their 4-momenta, event rate weights, weights for flux-averaged cross sections, weights for decay rates, and miscellaneous event information.
\darknus does not model detector geometries (except for its default implementation of MiniBooNE and MicroBooNE). 
For all other experiments, scattering on the list of desired targets is placed at $(t,x,y,z) = (0,0,0,0)$, and the decay position is calculated based on the propagating heavy neutrino.
All units in \darknus are in GeV for energies and momenta, cm for positions, and s for time.

\begin{figure*}[t]
    \centering
    \includegraphics[width=\textwidth]{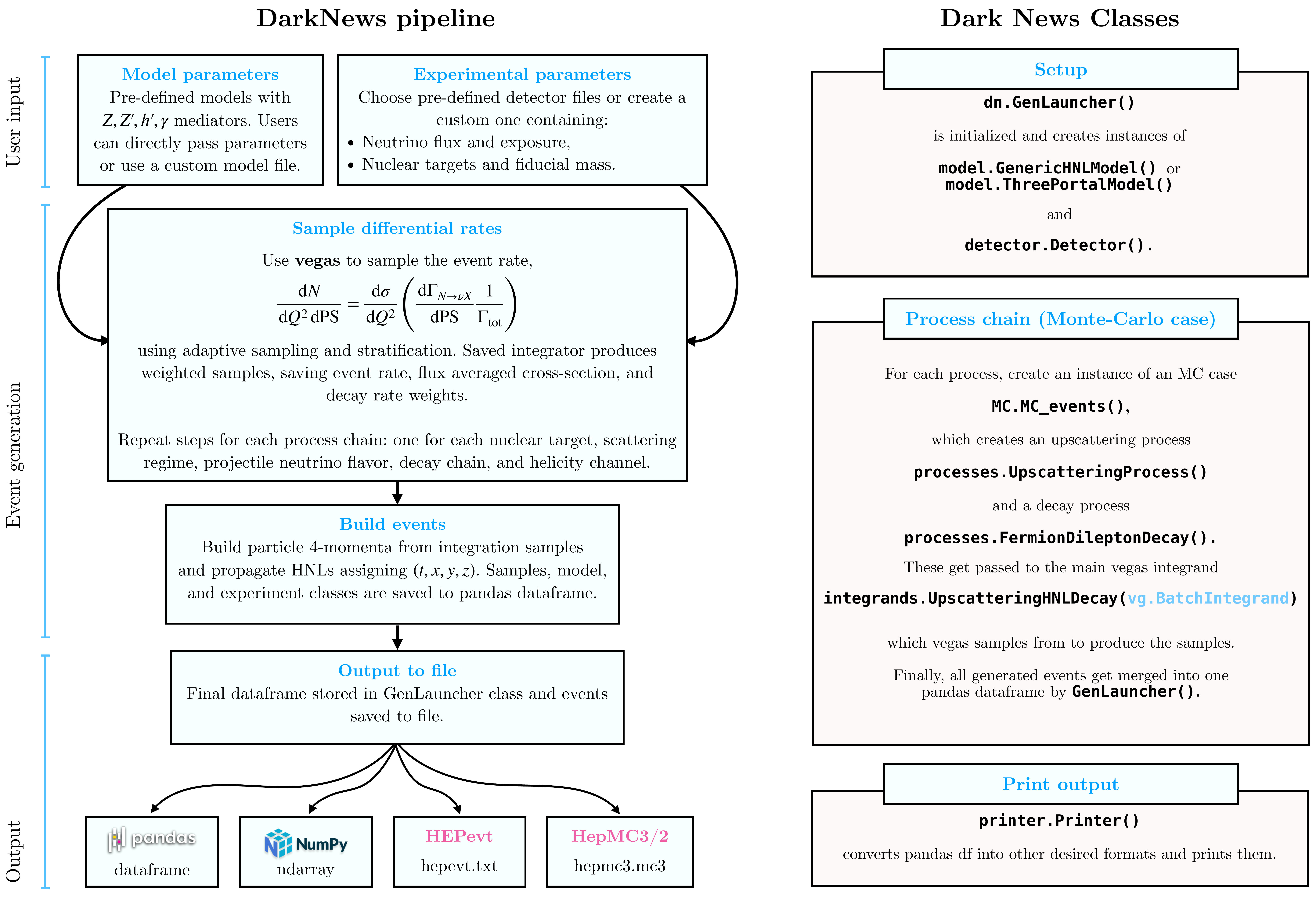}
    \caption{A schematic chart of the \darknus event generation. On the left, we provide a big-picture explanation of the generator in terms of the underlying physics, and on the right, in parallel, we show the \darknus classes responsible for each action.}
\end{figure*}
The code is publicly available on \textsc{g}it\textsc{h}ub~\footnote{\href{https://github.com/mhostert/DarkNews-generator}{{\large\color{blue-violet}\faGithub}\,\,{github.com/mhostert/DarkNews-generator}}.}, and can be used in a variety of accelerator neutrino experiments. 
\darknus was developed using \python~3.9 and makes use of a few \texttt{Cython} functions for improved performance.
As indicated in \Cref{fig:darknus_install}, the easiest way to install \darknus is via \texttt{pip}, which will also install all requisite requirements automatically.
Requirements include \numpy, \pandas, \vegas, and other smaller packages, including \texttt{particle} and \texttt{pyhepmc\_ng}, which are part scikit-hep~\cite{Rodrigues:2020syo}.
With the same \python environment used in the installation, \darknus can be easily imported in your \python script as \mintinline{python}|import DarkNews as dn|, defining our preferred shorthand \texttt{dn}.

Getting started with event generation is very simple.
We provide both a \python interface as well as a command-line functionality.
From a \python script, one can simply run
\begin{pythoncode}
import DarkNews as dn
my_gen = dn.GenLauncher(mzprime=1.25, m4=0.140, neval=1000, noHF=True, HNLtype="dirac", exp="microboone", nu_flavors=["nu_mu","nu_mu_bar"])
df = my_gen.run()
\end{pythoncode}
to generate approximately 1000 upscattering events per process chain (in this case 1000 for $\nu_\mu$ and 1000 for $\overline{\nu}_\mu$ initiated upscattering) at MicroBooNE for a dark photon mass of $1.25$~GeV and a Dirac heavy neutrino of mass $140$~MeV, excluding the sub-dominant, helicity-flipping (HF) chain.
The last line collects the main pandas dataframe with all the events and metadata, which is now also an attribute of the \mintinline{python}|GenLauncher| instance, \mintinline{python}|my_gen.df|.
The code above will also save the dataframe to a file at a default location set by the model parameter names.
Alternatively, it is possible to achieve the same result by using the command line,
\begin{shell-sessioncode}
dn_gen --mzprime 1.25 --m4 0.140 --neval 1000 --noHF --HNLtype dirac --exp microboone --nu_flavors nu_mu nu_mu_bar
\end{shell-sessioncode}
For single-valued arguments, an $=$ sign may be used, but for multi-valued arguments like in \texttt{--nu\_flavors} above, it cannot.

The two previous examples will use the default choices for couplings, but these can also be specified by the user, as we explain below.
A quick view into some of the kinematics and properties of the generation can be achieved with the \texttt{--make-summary-plots} option, which will produce a few histograms of energies and angles of the particles involved in the same path as the outputs are saved.

\vspace{-3ex}
\subsection{Inputs}

As shown above, it is possible to pass all model parameters as arguments to the \mintinline{python}|GenLauncher| class or to the \texttt{dn\_gen} command.
A list of available arguments is provided in \Cref{tab:input_parameters}. For the most up-to-date information, please see the \href{https://github.com/mhostert/DarkNews-generator/blob/main/README.md}{\texttt{README}} file.
The same list can also be accessed via \texttt{dn\_gen --help} in the terminal prompt.
Similarly, experimental definitions, which encompass a choice for neutrino flux, nuclear targets, and exposure, can also be set as arguments. 
In this case, the user can choose from a set of pre-defined experiments listed in \Cref{tab:exp_definitions}, together with our source for the neutrino fluxes used.
The given detector fiducial mass is an estimate that will vary from analysis to analysis.
Note that detector geometries are not specified within \darknus, and should be built by the user after event generation on the different nuclear targets is complete.
This means that the position of scattering is not implemented at the \darknus generation level (except for the MiniBooNE and MicroBooNE experiments) and that the decay position of the heavy neutrino is calculated from the origin at time $t=0$.

The user may define their own models and experimental configurations with input files. 
We provide two types of file interfaces for that, one for setting experimental conditions and one for setting model parameters.

\paragraph{Model definition files}
Input parameters can be specified via text file, which could be passed as a parameter to the \mintinline{python}|GenLauncher| class.
The following code would parse the parameters contained in the file \texttt{parameters.txt} (the search path is relative to the current working directory), loading their values prior to the arguments given to the \mintinline{python}|GenLauncher| constructor: in this example, the parameter \texttt{mzprime} is first read from the file \texttt{parameters.txt} and then from the parameter list.
\begin{pythoncode}
from DarkNews.GenLauncher import GenLauncher
gen_object = GenLauncher(param_file="parameters.txt", mzprime=1.4)
\end{pythoncode}
This is exactly equivalent to running the following line through the binary \texttt{dn\_gen}:
\begin{shell-sessioncode}
dn_gen --param-file parameters.txt --mzprime 1.4
\end{shell-sessioncode}

The \texttt{parameters.txt} file may look like the following.
\begin{center}
\begin{inputfile}[adjusted title=parameters.txt,width=0.9\linewidth]
hbar = 6.582119569e-25 # GeV s
c = 299792458.0 # m s^-1
gD = 2.0
alphaD = gD^2 / (4*PI)
epsilon = 1e-2
# chi = 0.0031
Umu5 = 1e-3
UD5 = 35.0
m4 = 0.080
m5 = 0.140
mzprime = 1.4
HNLtype = "majorana"
neval = 1000
\end{inputfile}
\end{center}

The following rules apply when writing an input parameter file:
\begin{enumerate}
    \item every line should be in the form of an assignment statement;
    \item a line is commented out by prepending it with \texttt{\#};
    \item the following input types are allowed:
    \begin{description}
        \item[integers] e.g. \texttt{5};
        \item[floats] written with decimals or in exponential notation, e.g. \texttt{5.0} or \texttt{5e10} (\texttt{5.} will result in an error);
        \item[strings] encapsulated with either single \texttt{`} or double \texttt{"} quotation marks, e.g. \texttt{"DarkNews"}, or \texttt{`DarkNews'}; escape character is the backslash \texttt{\textbackslash};
        \item[booleans] by explicitly writing \texttt{True} or \texttt{False}. These are case-insensitive;
        \item[lists] \python-like by encapsulating a comma-separated list of elements in square brackets; multi-line lists are also allowed;
    \end{description}
    \item it is possible to define any kind of variable, using also names not included among the allowed input parameters; variables can be defined according to the usual \python conventions for the variable name (only letters, digits and underscores) and with the assignment operator \texttt{=}. It is not allowed to name variables after program-defined names;
    \item inputs may be specified using mathematical expressions, according to the following rules:
    \begin{enumerate}[label=(\roman*)]
        \item \texttt{+}, \texttt{-}, \texttt{*}, \texttt{/} are the usual mathematical operators;
        \item \texttt{\^} is used for exponentiation (do not use \texttt{**});
        \item round brackets \texttt{(} and \texttt{)} can be used;
        \item \texttt{e} (case-insensitive, isolated: not inside float numbers) is understood as \python \mintinline{python}|math.e = 2.718281828|;
        \item \texttt{pi} (case-insensitive) is understood as \python \mintinline{python}|math.pi = 3.1415926535|;
        \item \texttt{sin(<expr>)}, \texttt{cos(<expr>)}, \texttt{tan(<expr>)} are the usual trigonometric functions;
        \item \texttt{exp(<expr>)} is the usual exponential function;
        \item \texttt{abs(<expr>)} is the absolute value;
        \item $\texttt{sgn(<expr>)} = -1$ if $\texttt{<expr>} < -10^{-100}$, $+1$ if $\texttt{<expr>} > 10^{100}$, $0$ otherwise;
        \item \texttt{trunc(<expr>)} returns the truncated float \texttt{<expr>} to integer;
        \item \texttt{round(<expr>)} is the integer part of the float number \texttt{<expr>};
        \item \texttt{sum(<list>)} will sum each element of the list \texttt{<list>}, returning a float number;
        \item any other string is interpreted as a variable which must already have been defined (the file is scanned from top to bottom).
    \end{enumerate}
\end{enumerate}
\paragraph{Experiment definition files}\label{sec:generator:input_exp}
\darknus provides the user with a range of pre-defined experiments, which may be accessed with a keyword through the \texttt{exp} input parameter:
\begin{pythoncode}
from DarkNews.GenLauncher import GenLauncher
gen_object = GenLauncher(exp="microboone")
\end{pythoncode}
equivalent to:
\begin{shell-sessioncode}
dn_gen --exp microboone 
\end{shell-sessioncode}
Keywords are listed along the experiments' features in \Cref{tab:exp_definitions}.

Alternatively, it is possible to define custom experiments via an experiment input file, e.g. \texttt{custom\_experiment.txt}, passed as a parameter to the \mintinline{python}|GenLauncher| class or to the binary \texttt{dn\_gen} in the same way as the parameter input file:
\begin{pythoncode}
from DarkNews.GenLauncher import GenLauncher
gen_object = GenLauncher(exp="custom_experiment.txt")
\end{pythoncode}
equivalent to:
\begin{shell-sessioncode}
dn_gen --exp custom_experiment.txt
\end{shell-sessioncode}
The search path of the file is also relative to the current working directory.
Notice that \darknus first looks for an experiment with a keyword equal to those defined in \Cref{tab:exp_definitions} to then afterward search for an experiment input file.
The experiment input file may look like the following:
\begin{center}
\begin{inputfile}[adjusted title=custom\_experiment.txt,width=0.91\linewidth]
name = "MicroBooNE"
fluxfile = "../fluxes/MicroBooNE_BNB_numu_flux.dat"
flux_norm = 1.0
erange = [
    0.05, 
    9
]
nuclear_targets = [
    'Ar40'
]
fiducial_mass = 85.0 # tons
fiducial_mass_per_target = [
    fiducial_mass
] 
POTs = 6.8e+20
\end{inputfile}
\end{center}
In this example, we define a custom MicroBooNE experiment that has a different number of POTs with respect to the default.
Notice that in this case, we wanted to keep the same definition for the neutrino fluxes file, so we defined the parameter \texttt{fluxfile} according to the path of the same file used by the default MicroBooNE experiment definition and listed in \Cref{tab:exp_definitions}.
In case we wanted to change the flux file, we could have set the path of a file relative to the directory where the experiment input file is stored.
Flux files are formatted as the following:
\texttt{Enu nue numu nutau nuebar numubar nutaubar}, the same one used in GLoBES~\cite{Huber:2004ka,Huber:2007ji}, where $E_\nu$ is in GeV and fluxes in $\nu/$cm$^2$/POT/GeV. In principle, they can have any units, which is normalized with the parameter \texttt{flux\_norm}.

A list of the possible parameters that can be set is shown in \Cref{tab:exp_parameters}.

\subsection{Output formats}

We now discuss the outputs of the event generation. 
Internally, \darknus uses a pandas dataframe containing all event information and generation setup.
The format is detailed in \Cref{tab:pandas_format}.

\renewcommand{\arraystretch}{1.3}
\begin{table*}[t]
    \centering
    \begin{tabular}{|>{\centering\arraybackslash}p{2.5 cm}|>{\centering\arraybackslash}p{1.5 cm}|>{\centering\arraybackslash}p{1.5 cm}|>{\centering\arraybackslash}p{2.3 cm}|>{\centering\arraybackslash}p{1.5 cm}|>{\centering\arraybackslash\tt}p{2.5 cm}|>{\centering\arraybackslash\tt}p{5cm}|}
        \hline 
        Detector & POT ($10^{20}$) & fiducial mass (t) & material & neutrino flux & {\normalfont keyword} & {\normalfont flux file}\\
        \hline\hline
        MiniBooNE FHC/RHC & $18.75$ ($11.27$) & $818$ & CH$_2$ & \cite{MiniBooNE:2008hfu} (data release) & miniboone\_fhc miniboone\_rhc & MiniBooNE\_FHC.dat MiniBooNE\_RHC.dat \\ \hline
        MicroBooNE & $12.25$ & $85$ & Ar & \cite{MiniBooNE:2008hfu} (data release) & microboone & MicroBooNE\_BNB\_numu\_flux.dat \\ \hline
        MINER$\nu$A LE FHC & $3.43$ & $6.10$ (in~\cite{MINERvA:2015nqi}) & C$_8$H$_8$ & \cite{MINERvA:2016iqn} (data release) & minerva\_le\_fhc & NUMI\_FHC\_LE.dat \\ \hline
        MINER$\nu$A ME FHC/RHC & $11.6$ ($12.2$) & $6.10$ (in~\cite{MINERvA:2019hhc}) & C$_8$H$_8$ & \cite{AliagaSoplin:2016shs} (digitized) & minerva\_me\_fhc minerva\_me\_rhc & NUMI\_FHC\_ME.dat (NUMI\_RHC\_ME.dat)\\ \hline
        NO$\nu$A ND LE & $13.6$ & $193$ (in~\cite{NOvA:2019bdw}) & H ($10.8 \%$), C ($66.7 \%$), O ($3.00 \%$), Cl ($16.1 \%$), Ti ($3.20 \%$) & \cite{nova_fluxes} (data release) & nova\_le\_fhc & NOvA\_FHC.dat \\ \hline
        MINOS LE & $10.56$ & $28.6$ (in~\cite{MINOS:2016yyz}) & C ($20\%$), Fe ($80\%$) & \cite{AliagaSoplin:2016shs} (digitized) & minos\_le\_fhc & NUMI\_FHC\_LE.dat \\ \hline
        T2K ND280 & $19.7$ & $22.34$ (in~\cite{Assylbekov:2011sh}) & H ($54 \%$), C ($3.6 \%$), \newline O ($1.8 \%$), Cu ($15 \%$),\newline Zn ($24 \%$), Pb ($1.9 \%$) & \cite{nd280_fluxes} (digitized) & nd280\_fhc & ND280\_FHC.dat \\ \hline
        DUNE LArND (1.2 MW, 120 GeV) & 100 & $30$ & Ar & \cite{DUNE:2021cuw} (data release) & dune\_nd\_fhc dune\_nd\_rhc & DUNE\_ND\_FHC.dat (DUNE\_ND\_RHC.dat)\\ \hline
        FASER$\nu$  &  units of $1/150$fb$^{-1}$ & 1.2 & W & \cite{Kling:2021gos} (data release) & fasernu & FASERnu.dat \\ \hline
        NuTeV & 0.05 & $690$ & C ($2 \%$), Fe ($98 \%$) & \cite{Avvakumov:2002cv} (digitized) & nutev\_fhc & NuTeV\_FHC.dat \\ \hline        
    \end{tabular}
    \caption{The default parameters assumed by \darknus for each different experiment. Note that the values above do not represent the full useful detector mass for each detector, which may be much larger depending on the analysis under consideration. 
    Flux files can be found in \texttt{"src/DarkNews/include/fluxes"} together with their original sources.
    }
    \label{tab:exp_definitions}
\end{table*}
\paragraph{Pandas dataframe} 
\darknus creates a pandas dataframe with all event information and 4-momenta built from the \vegas samples of each process chain. It then combines all of these into one final dataframe, which is pickled and saved to a file. 
It contains all the 4-momenta of the incoming and outgoing particles, including any intermediary particles, as well as weights and additional information about the events.
The dataframe index is a MultiIndex with columns and sub-columns.
That is to say, the dataframe is a 3-dimensional object with rows representing different events and columns representing an event property e.g. the 4-momenta of the particles, or the event rate weight. Sub-columns represent a Lorentz index (0,1,2,3) to denote $t,x,y,z$ coordinates, or $E,p_x,p_y,p_z$ momentum components.

Also included within the attributes of the pandas dataframe is the information used in the event generation. 
The attributes contain the model and detector classes used at the generation stage, the name of the generation (which can be specified by the user), and the lifetimes of all heavy neutrinos used.
The detector and model classes can be accessed via \mintinline{python}|df.attrs["model"]| and \mintinline{python}|df.attrs["experiment"]|, and contain all information regarding the model parameters, neutrino fluxes, target material, exposure, fiducial mass, etc.

\paragraph{sparse} It is also possible to remove less crucial information from the dataframe if one is not interested in non-observable particles or metadata, e.g. the four-momenta of invisible particles, or model parameters.
This ``sparse" dataframe can be useful when running a large number of simulations on a grid to save storage space.

\paragraph{Other \pandas/\numpy formats} Other output formats are also available. 
One example is the \texttt{.parquet} format, popular for its compression and fast read-and-write operations. 
The pandas dataframe saved to the parquet file is stripped of its \texttt{attrs} (no metadata).
Another possibility is to print all the floats in the dataframe to a \numpy array. This is not necessarily advantageous since pandas already work rather seamlessly with \numpy, but may be useful if one wants to avoid loading a pandas object when reading the event files. 
The column names are not included in the \texttt{.npy} files, so the user has to be careful to correctly identify the column indices.

\paragraph{HEPevt} Output formats that are more standard in particle physics can also be used.
The first one is \texttt{HEPevt}, a legacy \texttt{Pythia} format that is also used by several other generators.
There are two implementations of HEPevt formats: one legacy one that includes the event weight next to the event number and the number of particles in the event, and a more standard one implemented with the help of the scikit-hep package \texttt{pyhepmc}~\cite{pyhepmc-ng}. 
The latter format cannot be used with weighted events.

\paragraph{HepMC}
Finally, we also implement the more modern format \texttt{HepMC3}~\cite{Buckley:2019xhk,hepmc3} as well as the older \texttt{HepMC2} format.
These are also written using the \texttt{pyhepmc} package~\cite{pyhepmc-ng}.
This is a different \python binding for \texttt{HepMC3} from the one implemented in~\cite{hepmc3}, and allows us to stay fully inside \python, without the need for a local installation of \texttt{HepMC3}.
The event weights are saved in both cases with the same notation defined in \Cref{tab:pandas_format}.

\section{Models}\label{sec:models}

There are two model implementations in \darknus. The first follows a completely generic parametrization of the interaction vertices, called a \mintinline{python}|GenericHNLModel|. The second, and more physical, implementation follows the ``three-portal" model in Refs.~\cite{Ballett:2019pyw,Abdullahi:2020nyr} and is referred to as the \mintinline{python}|ThreePortalModel()|. 
The latter option is preferred when quoting results in terms of a physical and self-consistent model, but the former may be advantageous when the user is interested in implementing a model of their own.
For example, in the three-portal model, new scalar interactions between the heavy neutrinos and matter are rather suppressed and strongly constrained due to the $m_f/v_{\rm EW}$ suppression of the interactions.
In that case, non-minimal scalar sectors like those discussed in Refs~\cite{Dutta:2020scq,Datta:2020auq,Abdallah:2020biq,Abdallah:2020vgg,Abdallah:2022grs} can be implemented using the generic-model approach.

\begin{figure}[t]
    \centering
    \includegraphics[width=0.44\textwidth]{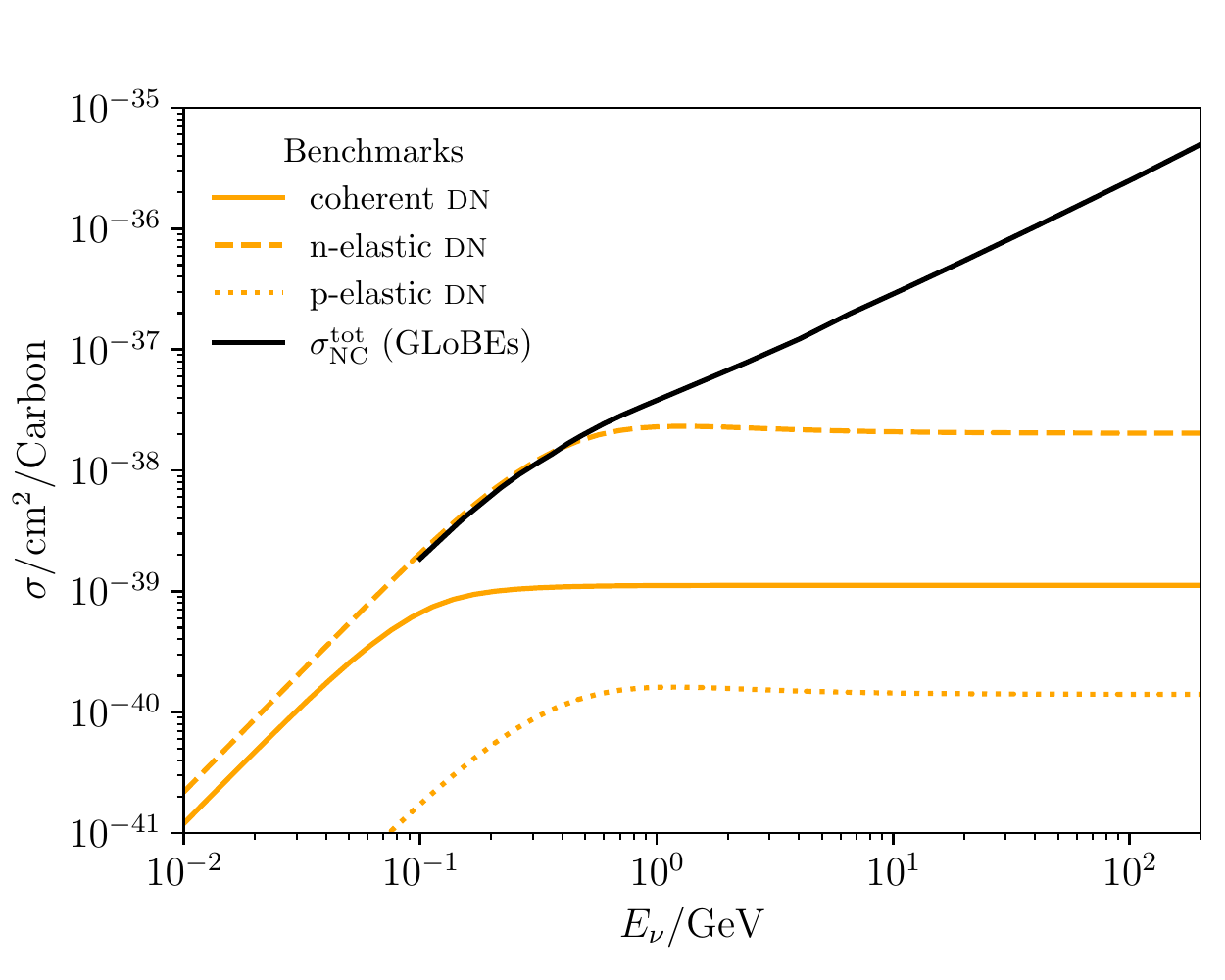}
    \caption{A comparison between the neutrino-neutron neutral current cross section calculation and the cross sections in Refs~\cite{Messier:1999kj,Paschos:2001np}. Neutron-elastic cross sections agree with the total neutral-current cross section up until the region where resonant and deep-inelastic-scattering regimes dominate.\label{fig:xsecNC}}
\end{figure}

\begin{figure}[t]
    \centering
    \includegraphics[width=0.49\textwidth]{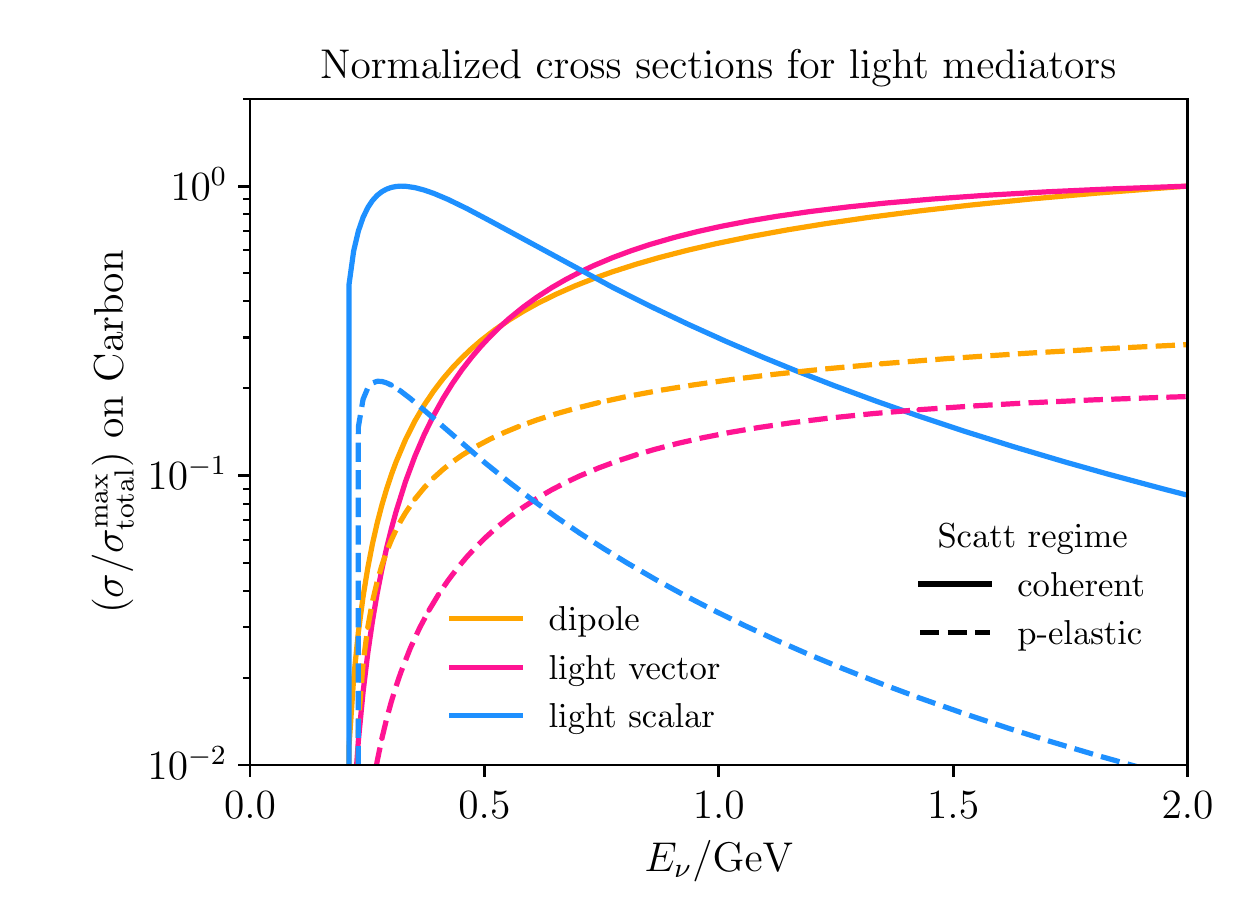}
    \includegraphics[width=0.49\textwidth]{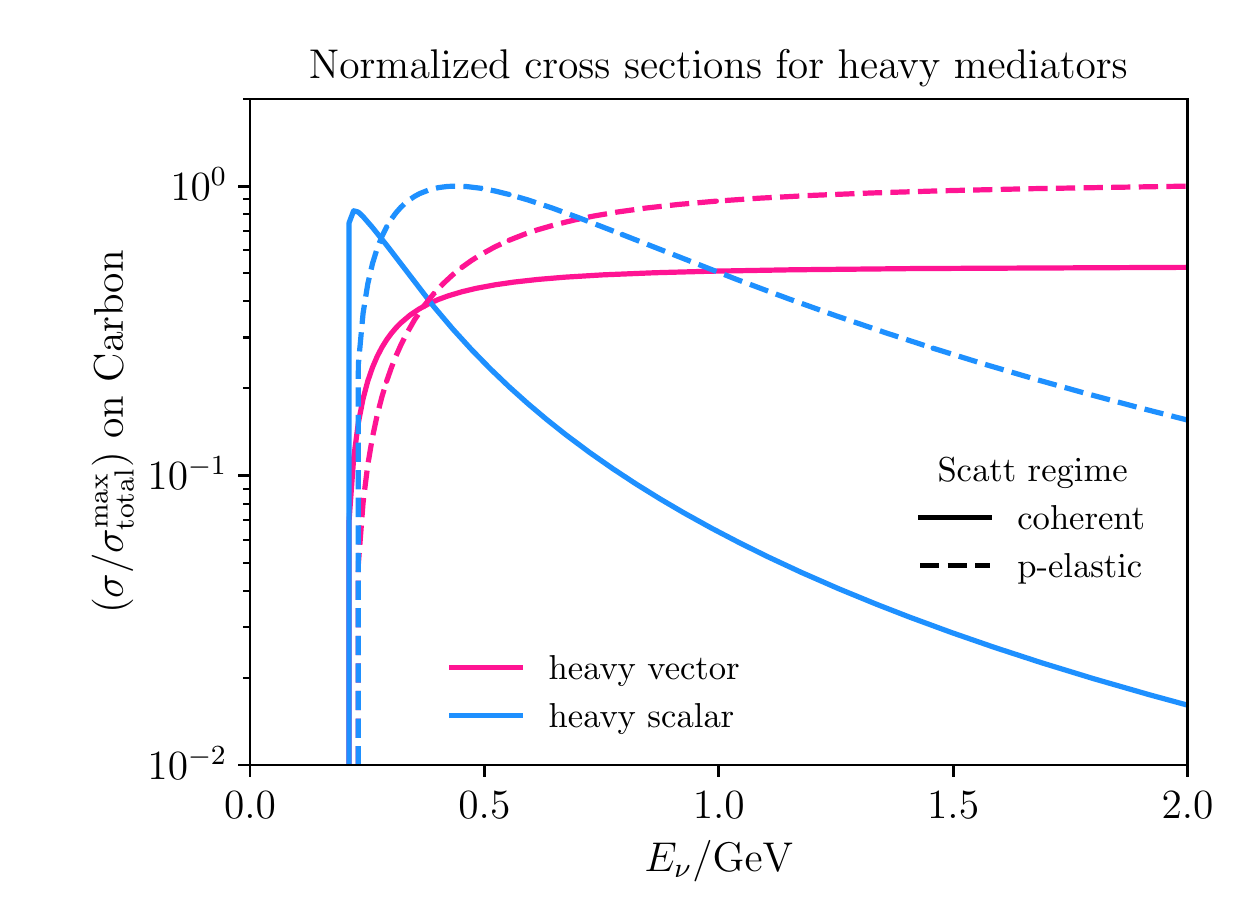}
    \caption{The shape of the total upscattering cross sections for different types of upscattering processes in \darknus, summed over final state helicities. 
    The top panel shows cases with light mediators ($m_{Z^\prime}=m_{h^\prime}=30$~MeV, including a transition magnetic moment), and the bottom one heavy mediators ($m_{Z^\prime}=m_{h^\prime}=1.25$~GeV). 
    \label{fig:xsecsDN}}
\end{figure}

\subsection{Model parameters}
\renewcommand{\arraystretch}{1.3}
\begin{table*}[t]
\centering
\begin{tabular}{|>{\centering\arraybackslash}p{1 cm}|>{\centering\arraybackslash}p{3.5 cm}|>{\centering\arraybackslash}p{7 cm}|>{\centering\arraybackslash}p{5.5 cm}|}
\hline
\multicolumn{1}{|c}{} &\multicolumn{1}{c|} {\textbf{ fermion flavor}} & \multicolumn{1}{c|} {\textbf{vector coupling}} & \multicolumn{1}{c|}{\textbf{axial-vector coupling}} \\
\hline
\hline
& & & \\
\multirow{9}{*}{$Z_\mu$} 
& charged leptons, $l^\pm$
& $c^l_V = \frac{g}{2 c_W} \left[ c_\beta \left(-\frac{1}{2}+2s_W^2\right) + \frac{3}{2}s_\beta s_W t_\chi\right]$ 
& $ c^l_A = \frac{g}{2 c_W} \left[-\frac{c_\beta + s_\beta s_W t_\chi}{2} \right]$ \\
& & & \\
& up-type quarks, $q^u$
& $ c^u_V = \frac{g}{2 c_W} \left[c_\beta \left(\frac{1}{2}-\frac{4}{3}s_W^2\right) - \frac{5}{6}s_\beta s_W t_\chi\right]$ 
& $c^u_A = \frac{g}{2 c_W} \left[\frac{c_\beta + s_\beta s_W t_\chi}{2}\right]$ \\ 
& & & \\
& down-type quarks, $q^d$
& $(c^d_{ij})_V = \frac{g}{2 c_W} \left[c_\beta \left(-\frac{1}{2}+\frac{2}{3}s_W^2\right) + \frac{1}{6} s_\beta s_W t_\chi\right]$ 
& $(c^d_{ij})_A = \frac{g}{2 c_W} \left[-\frac{c_\beta + s_\beta s_W t_\chi}{2}\right]$ \\
& & & \\
& \multirow{3}{*}{heavy neutral lepton, N}
& $\begin{aligned}[t] c^{N}_V = \frac{g}{4c_W}U^\dagger_{\alpha i}&U_{\alpha j}\left(c_\beta + s_\beta s_W t_\chi\right)  \\ + &\frac{s_\beta \, g_D }{2c_\chi} U^\dagger_{D i}U_{D j} \end{aligned}$ & \multirow{3}{*}{$c^{N}_A = -c^{N}_V$}\\
& & & \\
\hline
\hline
& & & \\
\multirow{ 9}{*}{$Z^\prime_\mu$} 
& charged leptons, $l^\pm$
& $d^l_V = \frac{g}{2 c_W} \left[ c_\beta \left(-\frac{1}{2}+2s_W^2\right) + \frac{3}{2}s_\beta s_W t_\chi\right]$ 
& $ d^l_A = \frac{g}{2 c_W} \left[-\frac{c_\beta + s_\beta s_W t_\chi}{2} \right]$ \\ 
& & & \\
& up-type quarks, $q^u$
& $ d^u_V = \frac{g}{2 c_W} \left[c_\beta \left(\frac{1}{2}-\frac{4}{3}s_W^2\right) - \frac{5}{6}s_\beta s_W t_\chi\right]$ 
& $d^u_A = \frac{g}{2 c_W} \left[\frac{c_\beta + s_\beta s_W t_\chi}{2}\right]$\\ 
& & & \\
& down-type quarks, $q^d$
& $d^d_V = \frac{g}{2 c_W} \left[c_\beta \left(-\frac{1}{2}+\frac{2}{3}s_W^2\right) + \frac{1}{6} s_\beta s_W t_\chi\right]$ 
& $d^d_A = \frac{g}{2 c_W} \left[-\frac{c_\beta + s_\beta s_W t_\chi}{2}\right]$ \\
& & & \\
&  \multirow{3}{*}{heavy neutral lepton, $N$}
& $\begin{aligned}[t] (d^{N}_{ij})_V =  - \frac{g}{4c_W}U^\dagger_{\alpha i}&U_{\alpha j}\left(s_\beta - c_\beta s_W t_\chi\right) \\
+ & \frac{c_\beta \, g_D}{2c_\chi} U^\dagger_{D i}U_{D j} \end{aligned}$
& \multirow{3}{*}{$(d^{N}_{ij})_A = -d^{N}_V$} \\
& & & \\
\hline
\end{tabular}
\caption{Vector and axial-vector couplings of the SM charged fermions and heavy neutral leptons (which include the SM light neutrinos) with the modified $Z_\mu$ and new $Z^\prime_\mu$ gauge bosons. 
Note that $c_x = \cos x$ and $s_x = \sin x$, where $s_W$ is the sine of the Weinberg angle, $\tan{2\beta} \equiv (2 s_W s_\chi c_\chi)/(c_\chi^2 - s_W^2 s_\chi^2 - \mu^2)$, $\mu=m_{Z^\prime}/m_Z$, and $\chi = \epsilon/c_W$. In these expressions, we have assumed that dark neutrinos have a unit charge under the dark gauge group.
\label{tab:portal_couplings}}
\end{table*}

\paragraph{Generic model} To assign interaction vertices, this model definition makes use of the following simplified interactions terms,
\begin{align}\label{eq:Ninteractions}
    \mathcal{L}_{\rm neutral} &\supset c_{ij} \,Z_\mu \overline{N_i} \gamma^\mu N_j + d_{ij} \,Z^\prime_\mu \overline{N_i} \gamma^\mu N_j    \\ \nonumber
    & \quad + s_{ij} \, h^\prime \overline{N_i} N_j + \frac{\mu^{\rm tr}_{ij}}{2} \overline{N_i} \sigma_{\mu\nu}F^{\mu\nu} N_j,
\end{align}
where $N_i$ is some neutral lepton, light or heavy, in the set $\{\nu_e, \nu_\mu, \nu_\tau, N_4, N_5, N_6\}$. 
Note that for light neutrinos, we work with the flavor index, rather than the mass index. 
Here, $c_{ij}, d_{ij}, s_{ij}$ and $\mu^{\rm tr}_{ij}$ parametrize the couplings of the neutral leptons with the SM $Z$ boson, the dark photon, the dark scalar, and transition magnetic moment terms.
The mediators are also allowed to have couplings to other SM particles. 
For the couplings to charged particles, we include
\begin{align}\label{eq:SMinteractions}
    \mathcal{L}_{\rm charged} &\supset 
    Z_\mu \bar{f} \left( c^{f}_V \gamma^\mu + c^{f}_A \gamma^\mu\gamma^5 \right) f
    \\\nonumber&+
    Z^\prime_\mu \bar{f} \left( d^{f}_V \gamma^\mu + d^{f}_A \gamma^\mu\gamma^5 \right) f
    \\\nonumber&+
    h^\prime \bar{f} \left( d^{f}_S + d^{f}_P \gamma^5 \right) f,
\end{align}
where $f = \{u,d,e\}$. 
It is always assumed that the direct couplings to electrons and muons are the same.

 \paragraph{Three-portal model} It is also possible to set the dark photon and dark scalar interaction vertices using the three portal model of Refs.~\cite{Ballett:2019pyw,Abdullahi:2020nyr}, which features a dark neutrino, $\nu_D$, in addition to the standard sterile neutrino, $\nu_N$, and the dark photon and dark scalar mediators.
 In this case, the interaction vertices defined in \cref{eq:SMinteractions} may be computed with the parameters, $U_{\alpha i}, \, U_{D i}, \, \epsilon, \, g_D, \text{ and } \theta$, as featured in the code. Here, $U_{\alpha i}$ refers to the neutrino mixing matrix elements in the light neutrino sub-block, $U_{D i}$ to the elements in the dark neutrino sub-block, $\epsilon$ to the kinetic mixing parameter, $g_D$ to the coupling of the dark photon to the dark neutrino, and $\theta$ to the mixing between the SM Higgs and new dark scalar mediator.

\begin{figure*}[t]
    \centering
    \includegraphics[width=0.32\textwidth]{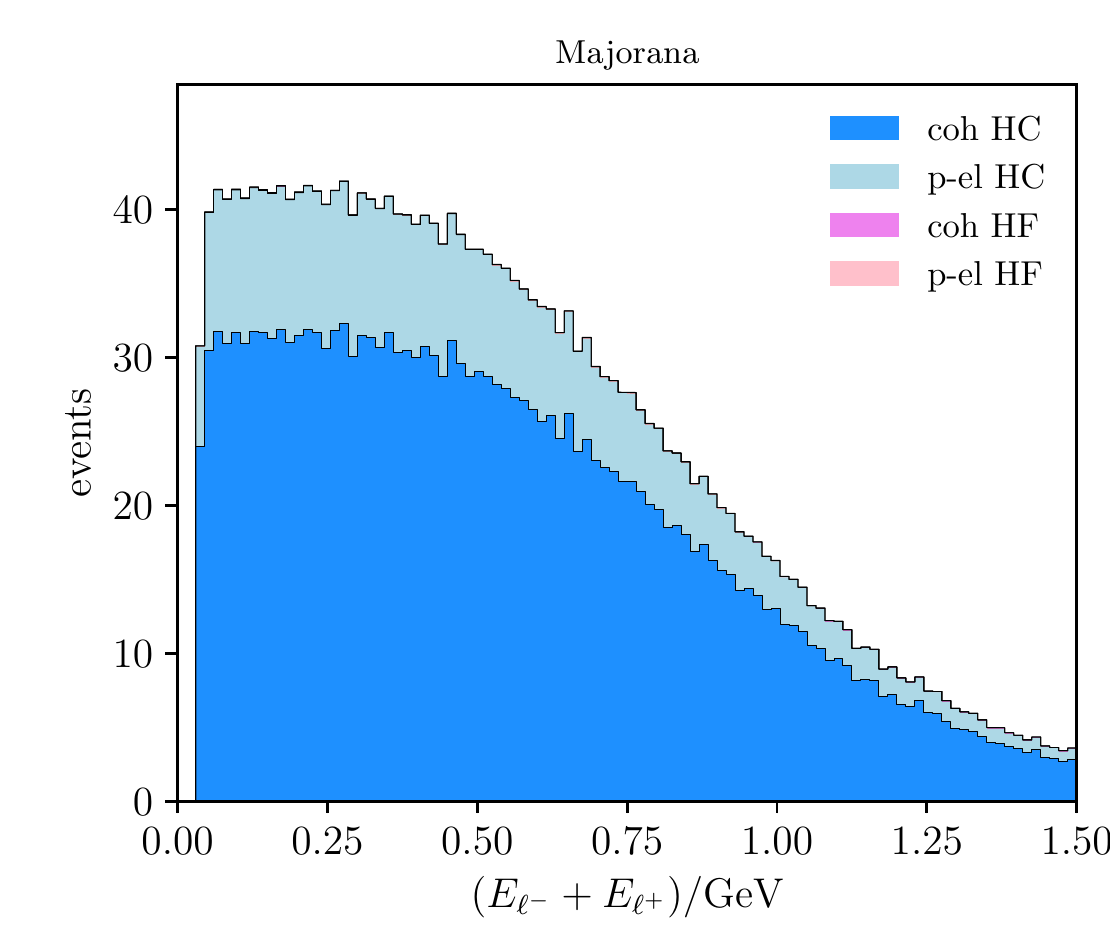}
    \includegraphics[width=0.32\textwidth]{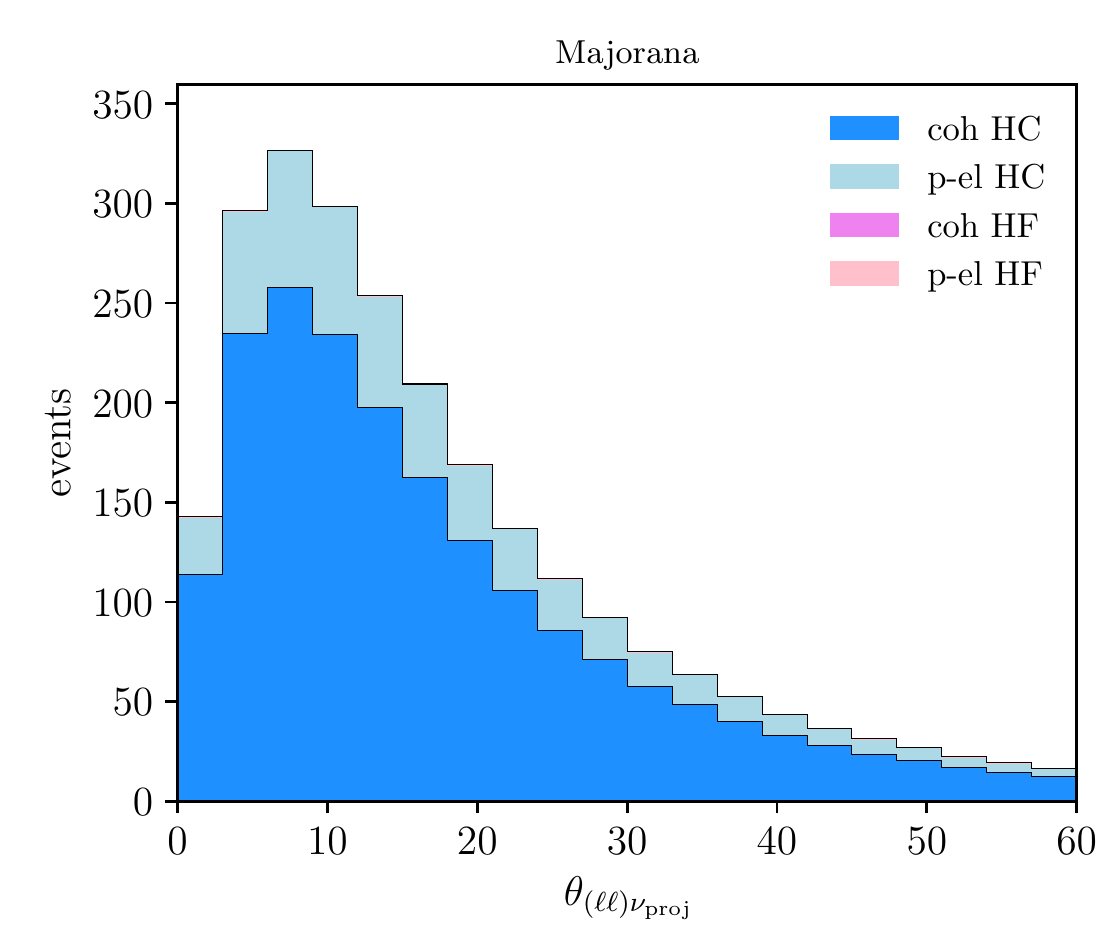}
    \includegraphics[width=0.32\textwidth]{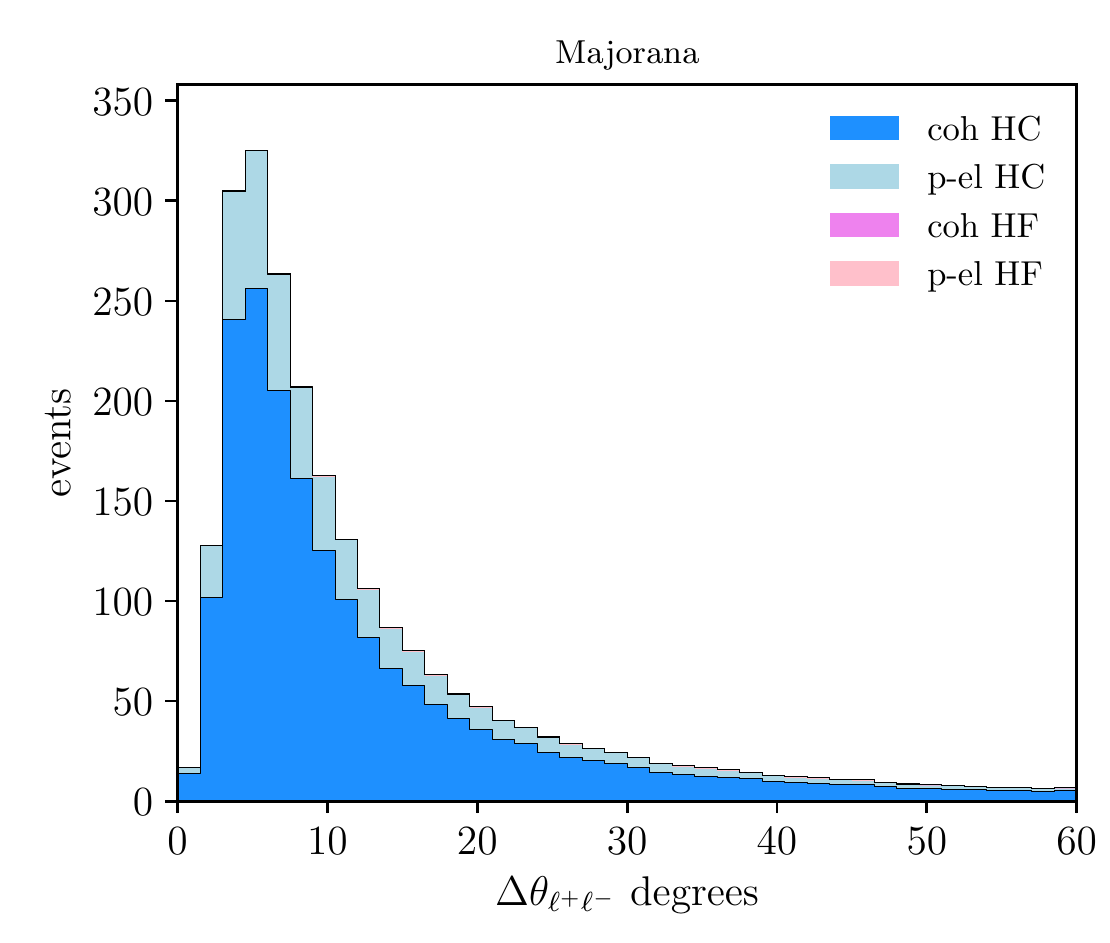}
    \\
    \includegraphics[width=0.32\textwidth]{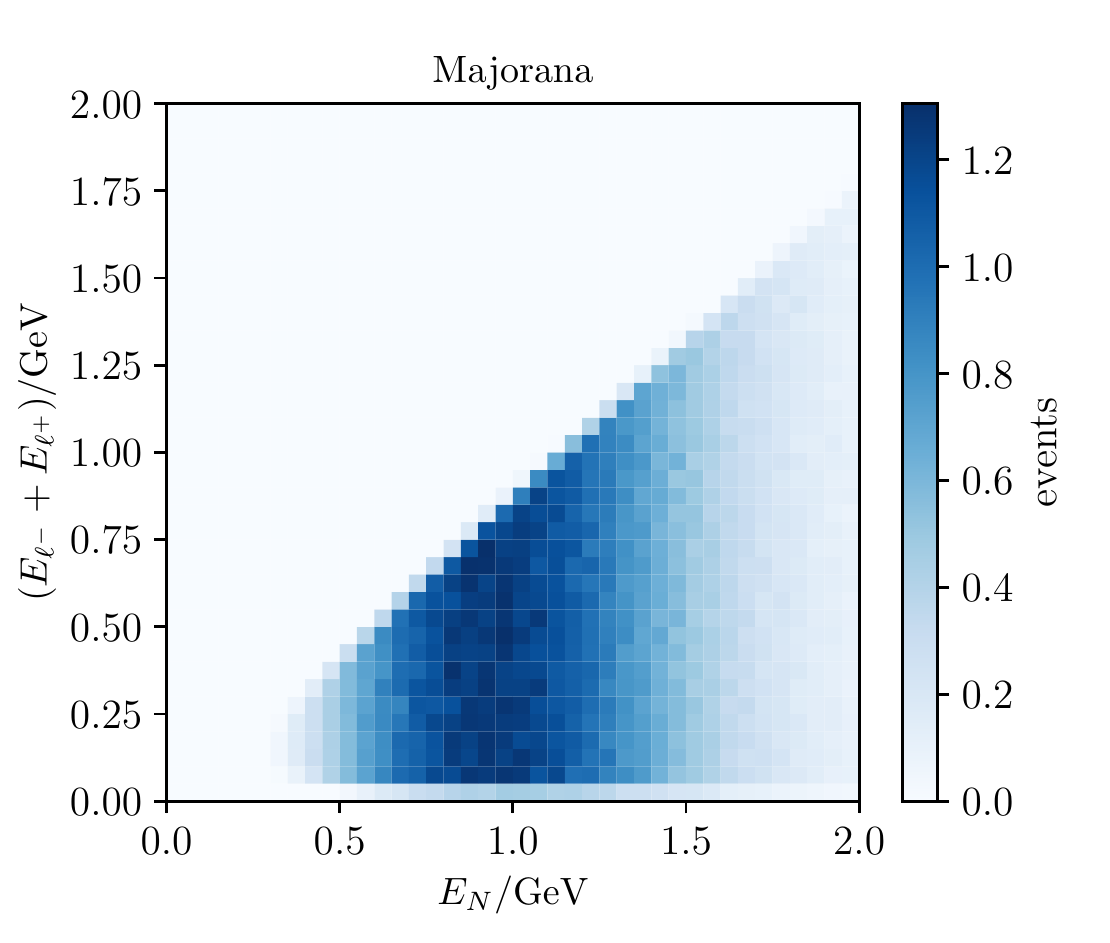}
    \includegraphics[width=0.32\textwidth]{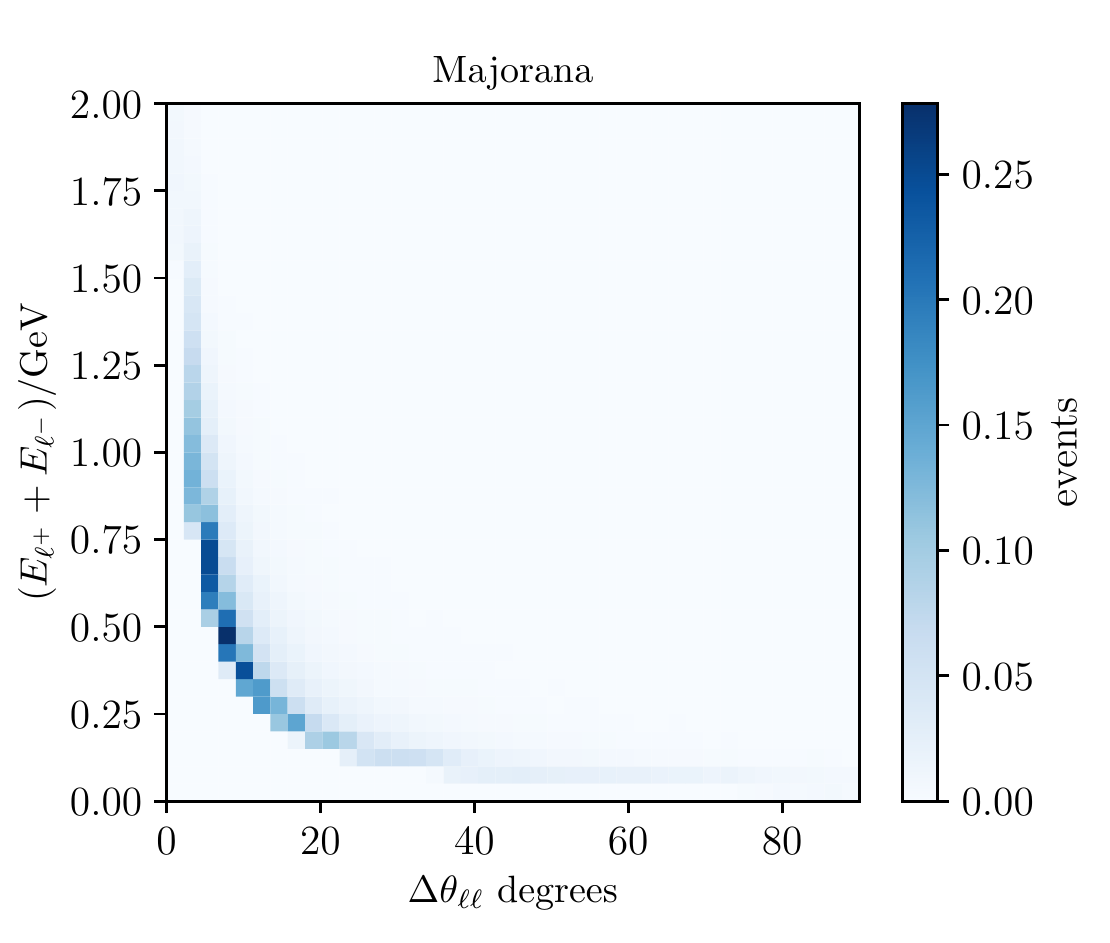}
    \includegraphics[width=0.32\textwidth]{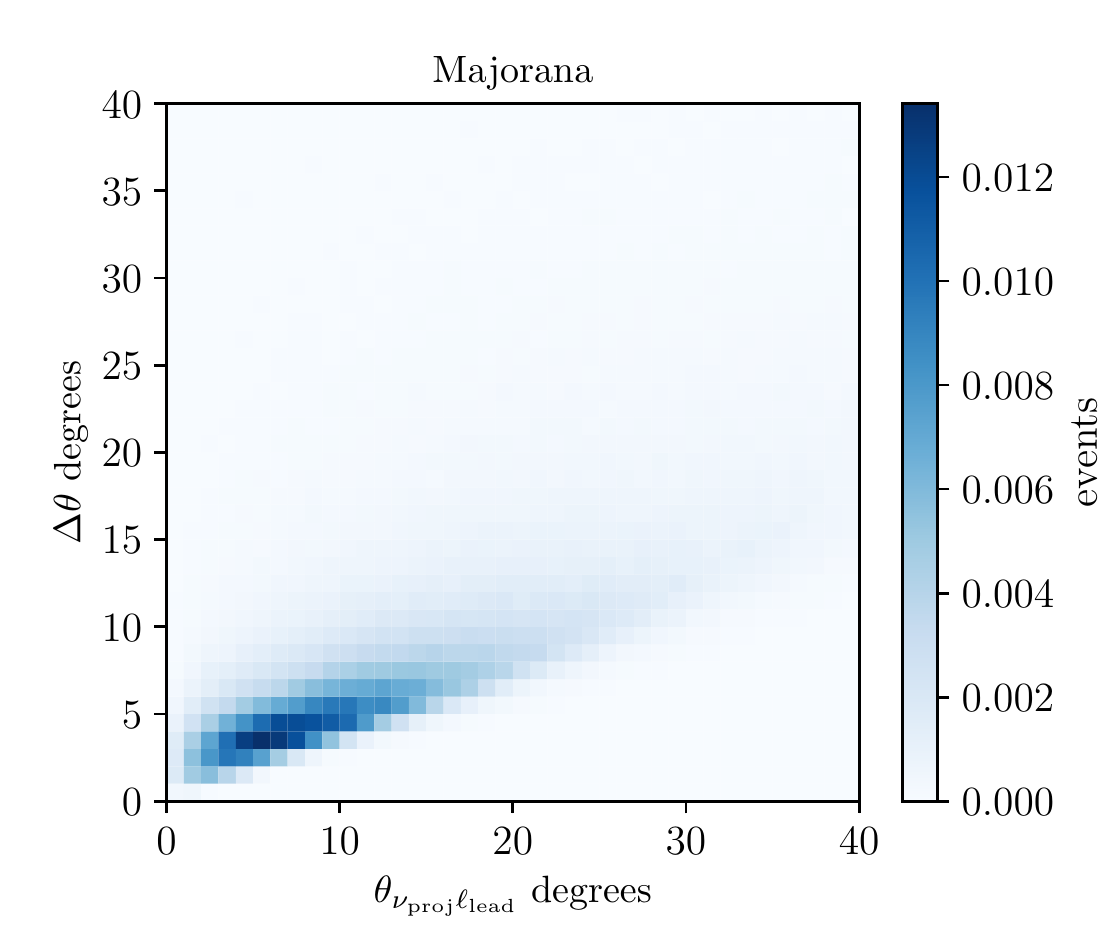}
    \caption{Stacked histograms of the kinematical distributions of the final states produced in the neutrino-nucleus upscattering. Heavy neutrino is a Majorana particle with model parameters taken to be: $g_D = 1$, $\epsilon = 10^{-4}$, $m_5 = 300$~MeV,  $m_4 = 100$~MeV, and  $m_{Z^\prime} = 30$~MeV. See text for definitions.\label{fig:distributions_zprime_maj}}
\end{figure*}
In the language of the couplings introduced in the generic model, the full expressions for the couplings in the three-portal model are given in \Cref{tab:portal_couplings} (see also Ref.~\cite{Babu:1997st}).
The scalar couplings to heavy neutrinos is always set in a model-independent way,
as it would otherwise be necessary to specify the Yukawa couplings and mixing elements between the sterile and the heavy neutrino, $U_{Ni}$.
This approach is justified given that for most applications, interesting phenomenology can only be achieved in non-minimal scalar models for which the generic model approach is more appropriate.

In the limit of small portal couplings, we have that
\begin{align}
    \mathcal{L}_{\rm charged} \supset e\epsilon q_f \, Z^\prime_\mu \bar{f} \gamma^\mu f + \theta \frac{m_f}{v_{\rm EW}}h^\prime \bar{f} f,
\end{align}
where $e$ is the electromagnetic (EM) coupling, $q_f$ the EM charge of the fermion $f$, $g$ the electroweak gauge coupling, 
$c_W$ the cosine of the Weinberg angle, $m_f$ the mass of the fermion $f$, and $v_{\rm EW}$ the Higgs vev. 
Scalar-nucleon interactions are described according to Ref.~\cite{Cline:2013gha}.
In this model, charged-current diagrams in the HNL decays are also taken into account, and scale proportionally to $U_{\alpha i}$.

\subsection{Cross sections}

We now discuss how the analytical expressions for the upscattering cross sections are calculated.
First, let us consider the process of a normalized low-energy neutrino flavor state 
\begin{equation}
\ket{\hat{\nu}_{\alpha}} = \frac{ \sum_{i=1}^{3} U_{\alpha i} \ket{\nu_i}}{\sqrt{\sum_i^{3}|U_{\alpha i}|^2}}
\end{equation}
scattering on a hadronic target $\mathcal{H}$, either a nucleus or free nucleon, producing a heavy neutral lepton $N_j$ with helicity $h$,
\begin{equation}
    \hat{\nu}_\alpha(k) + \mathcal{H}(p) \to N_j^{h=\pm1} (k^\prime) + \mathcal{H}(p+q),
\end{equation}
where $q = k- k^\prime = (\nu, \vec{q})$ is the momentum transfer to the hadronic target. 
We define $\nu = E_\nu - E_N$ and $Q^2 \equiv - q^2$.

For convenience of notation and to keep the connection to the indexing used in the code, we will assign $i=\{e, \mu, \tau\}\equiv \{1,2,3\}$.
This effectively means that we are considering a diagonal mass matrix for the three light neutrinos,
and assuming that heavy neutrinos with $i\geq 4$ always decohere from the light states.
\begin{figure*}[t]
    \centering
    \includegraphics[width=0.32\textwidth]{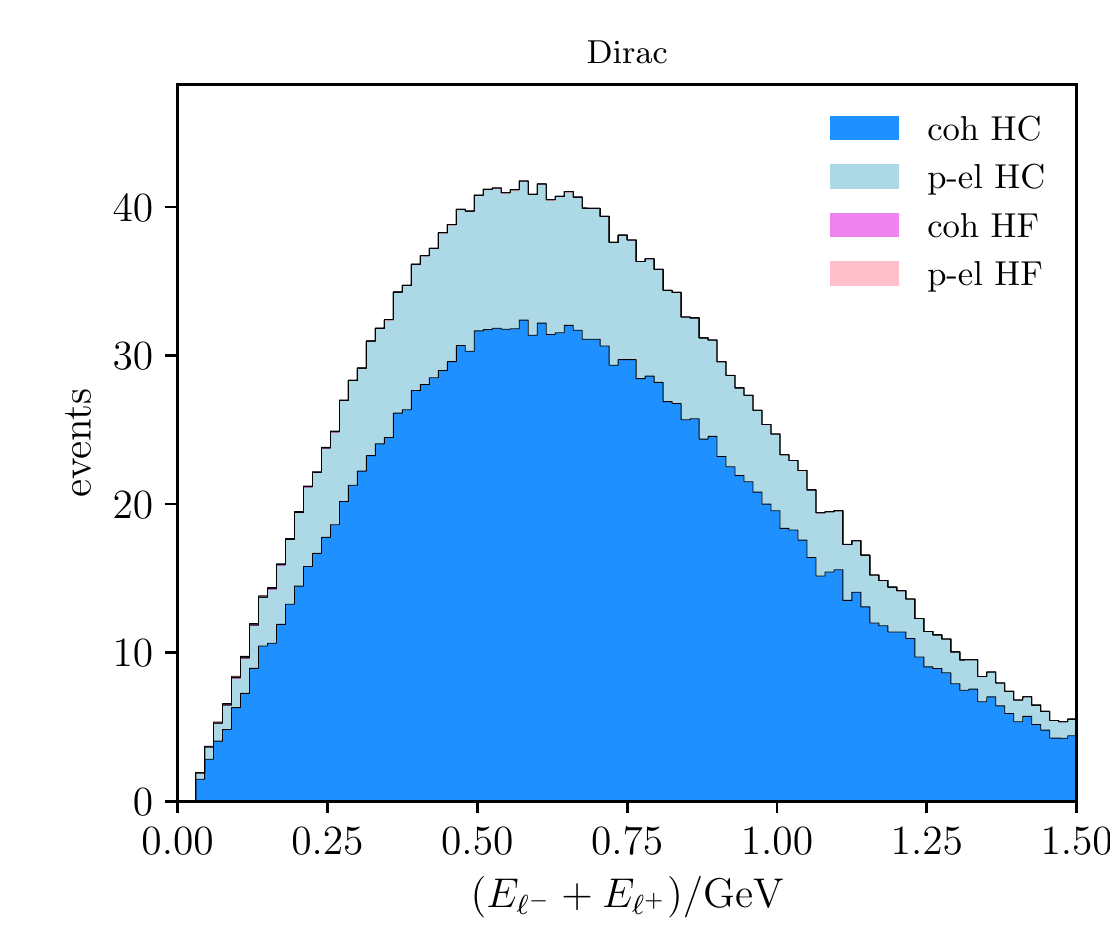}
    \includegraphics[width=0.32\textwidth]{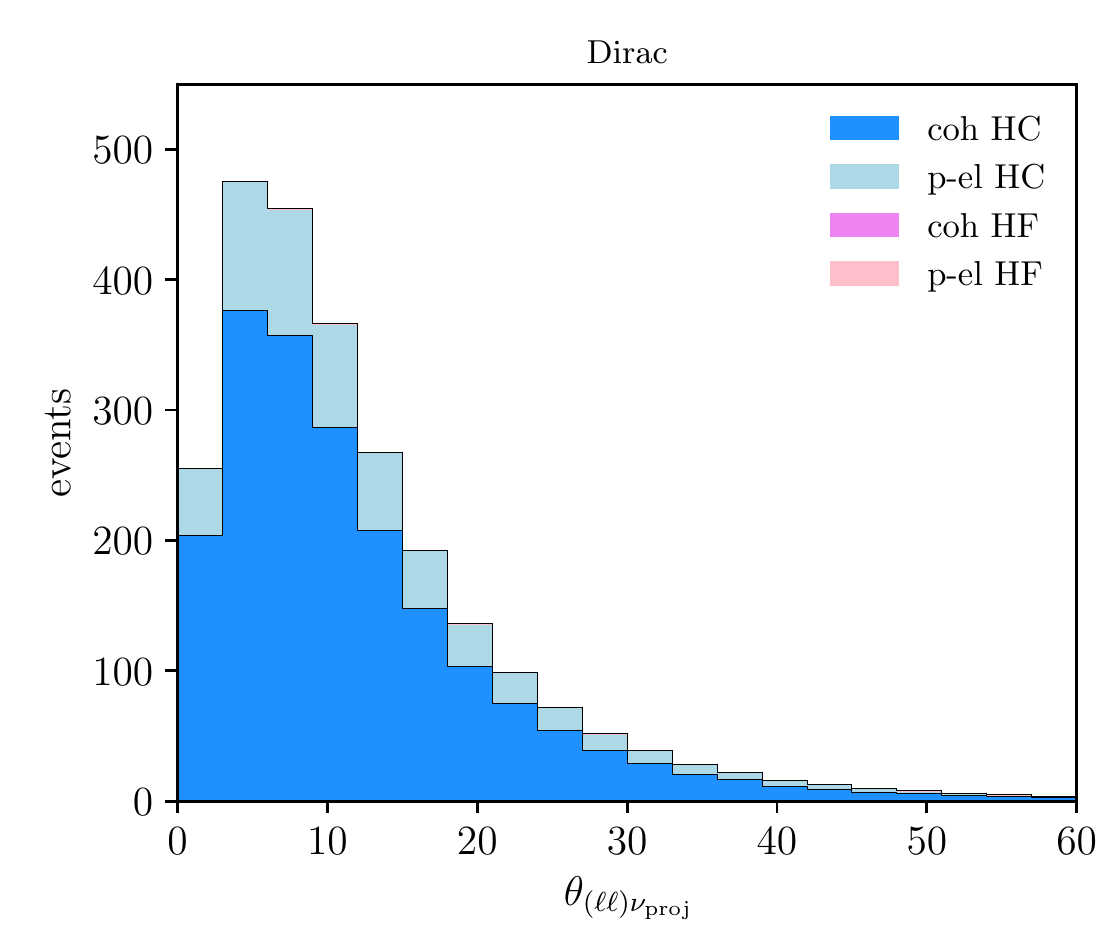}
    \includegraphics[width=0.32\textwidth]{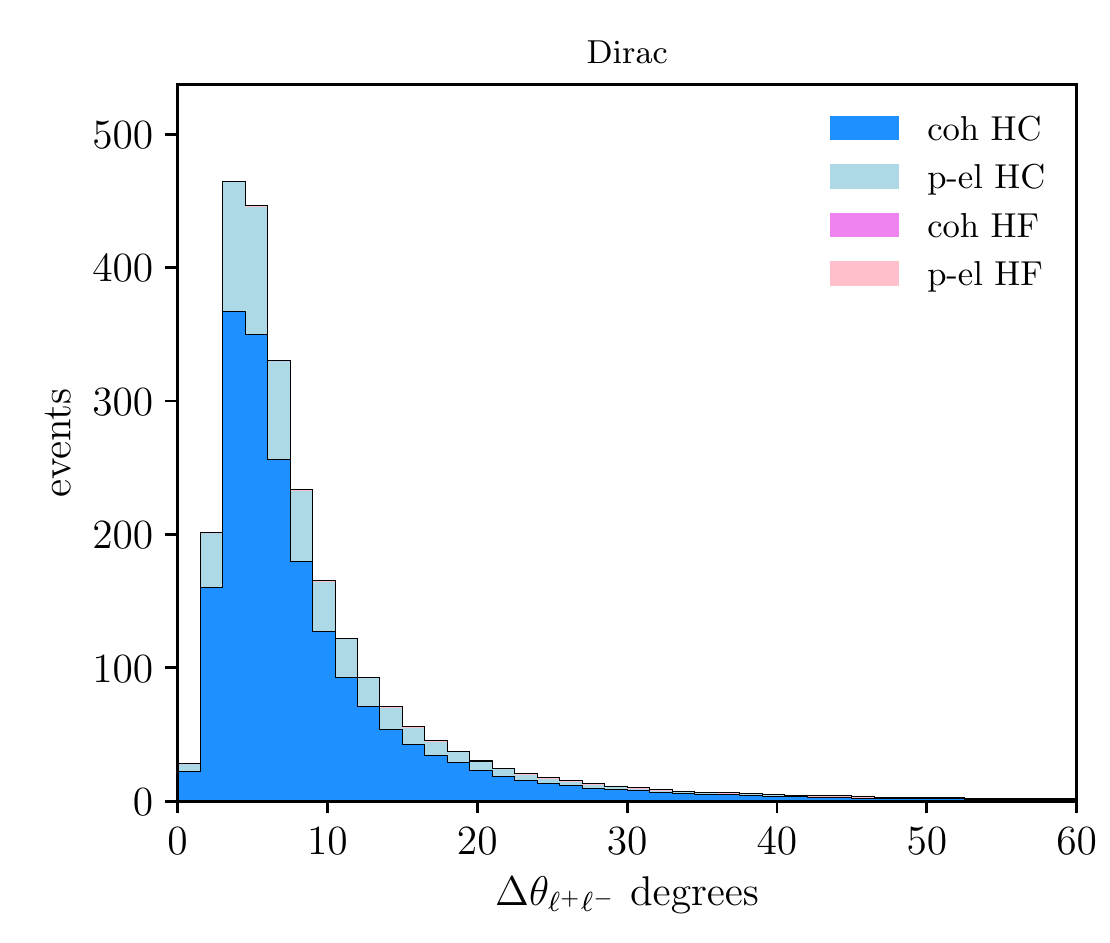}
    \\
    \includegraphics[width=0.32\textwidth]{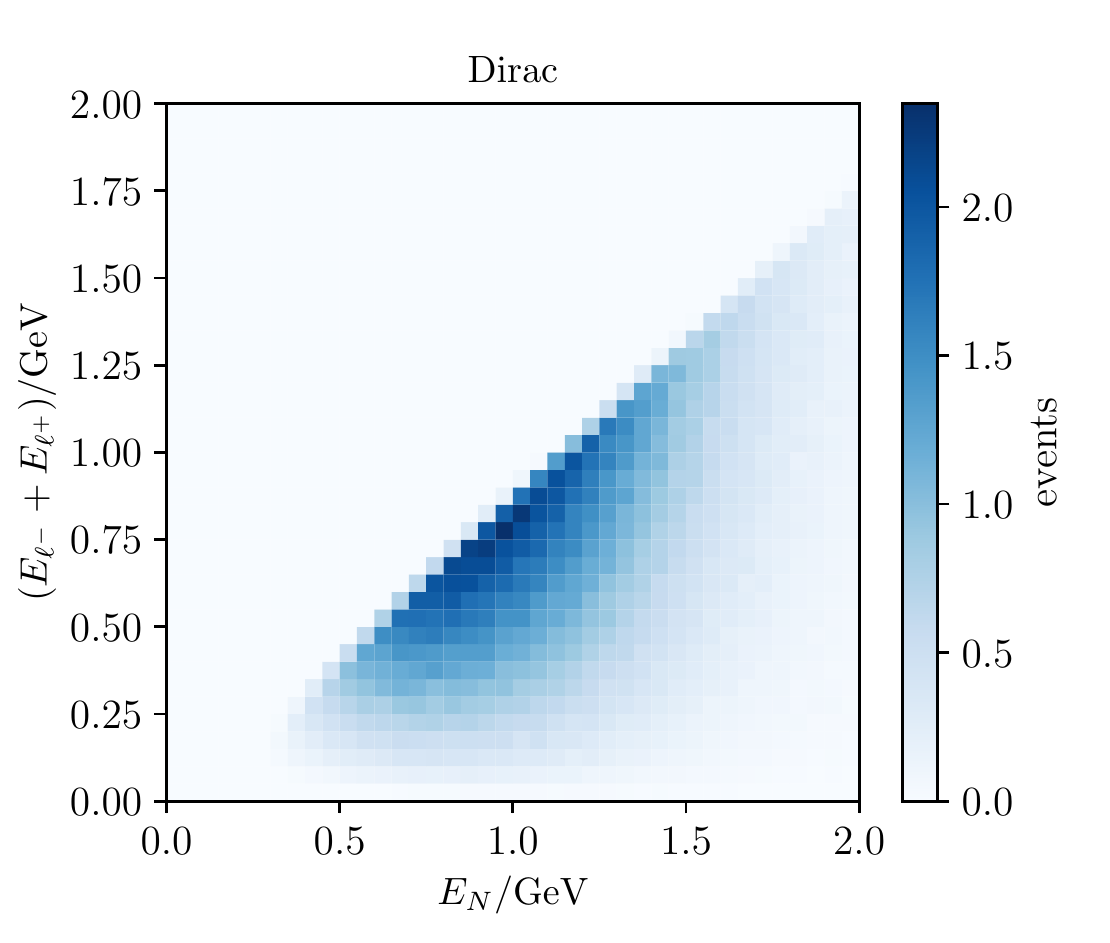}
    \includegraphics[width=0.32\textwidth]{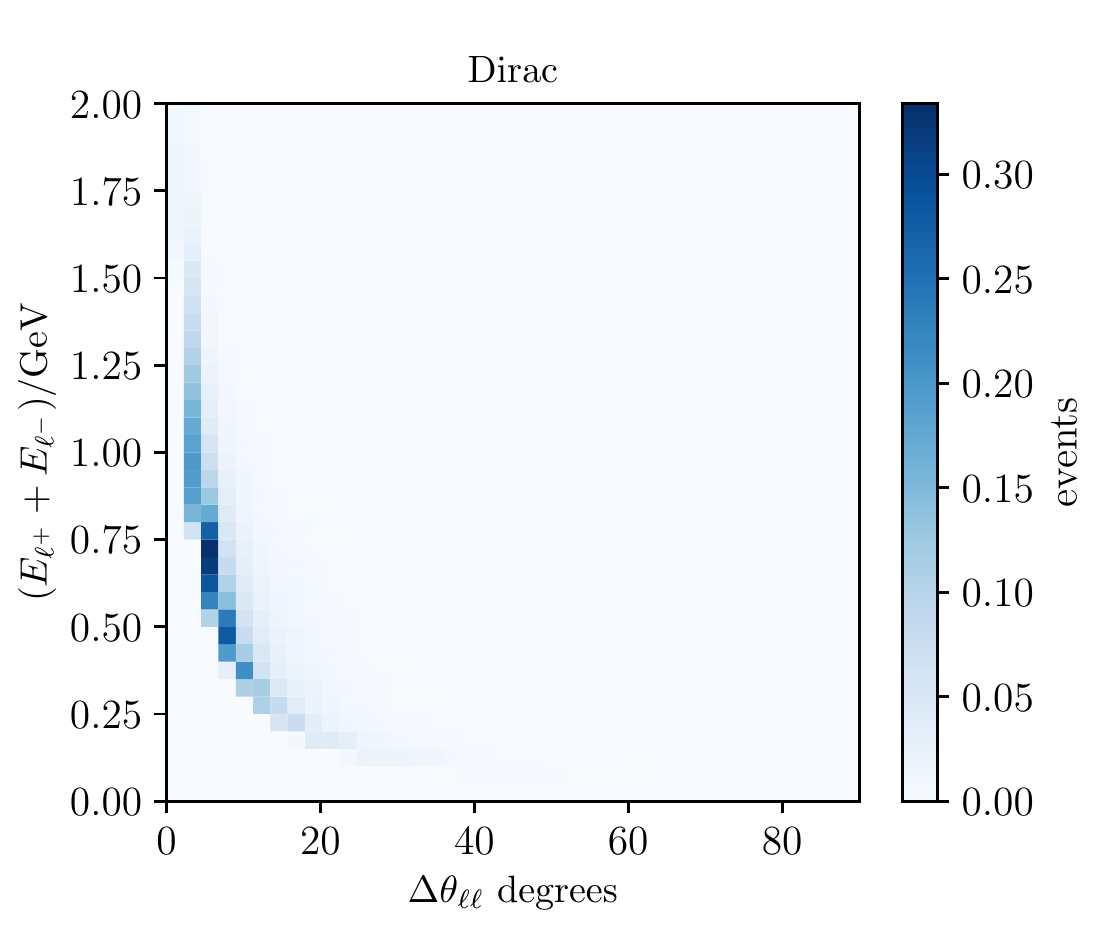}
    \includegraphics[width=0.32\textwidth]{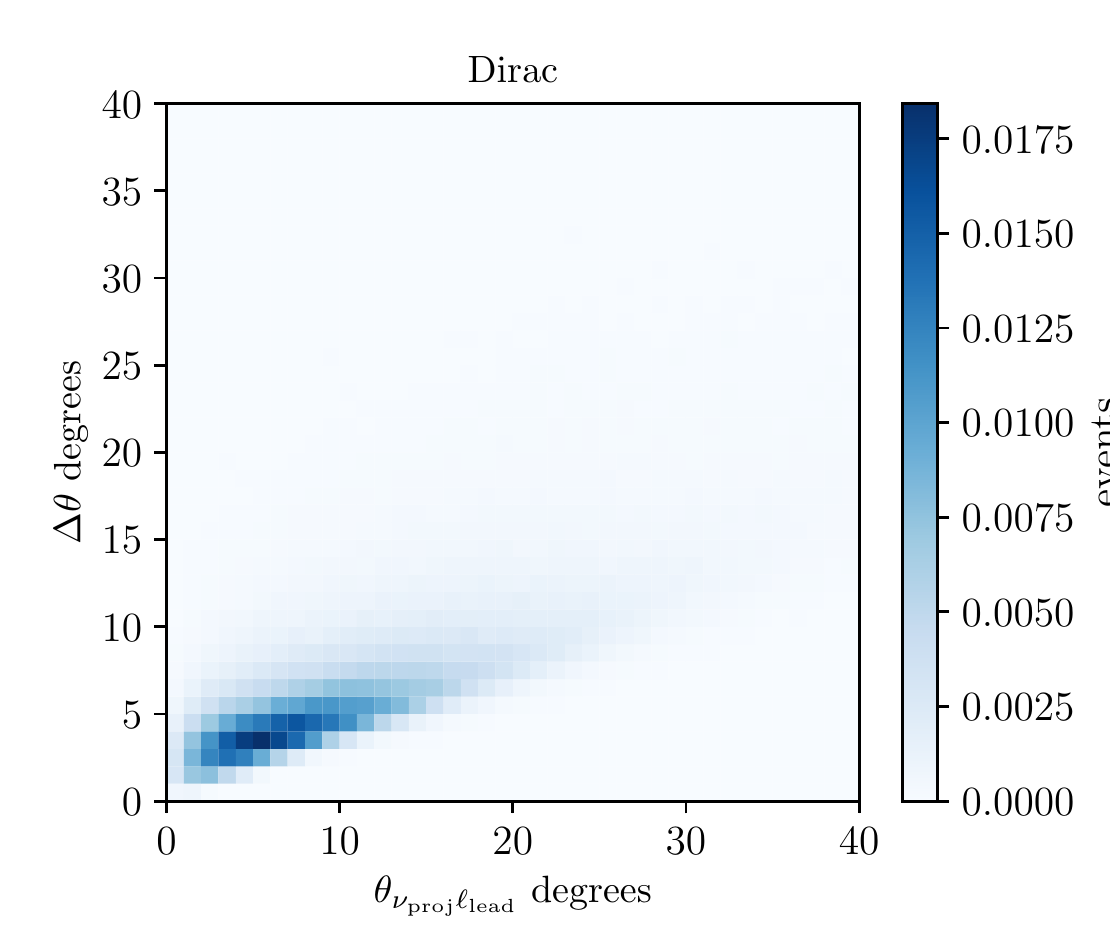}
    \caption{Stacked histograms of the kinematical distributions of the final states produced in the neutrino-nucleus upscattering. Heavy neutrino is a Dirac particle with model parameters taken to be: $g_D = 1$, $\epsilon = 10^{-4}$, $m_5 = 300$~MeV,  $m_4 = 100$~MeV, and  $m_{Z^\prime} = 30$~MeV. See  text for definitions.
    \label{fig:distributions_zprime_dir}}
\end{figure*}

\begin{figure*}[t]
    \centering
    \includegraphics[width=0.32\textwidth]{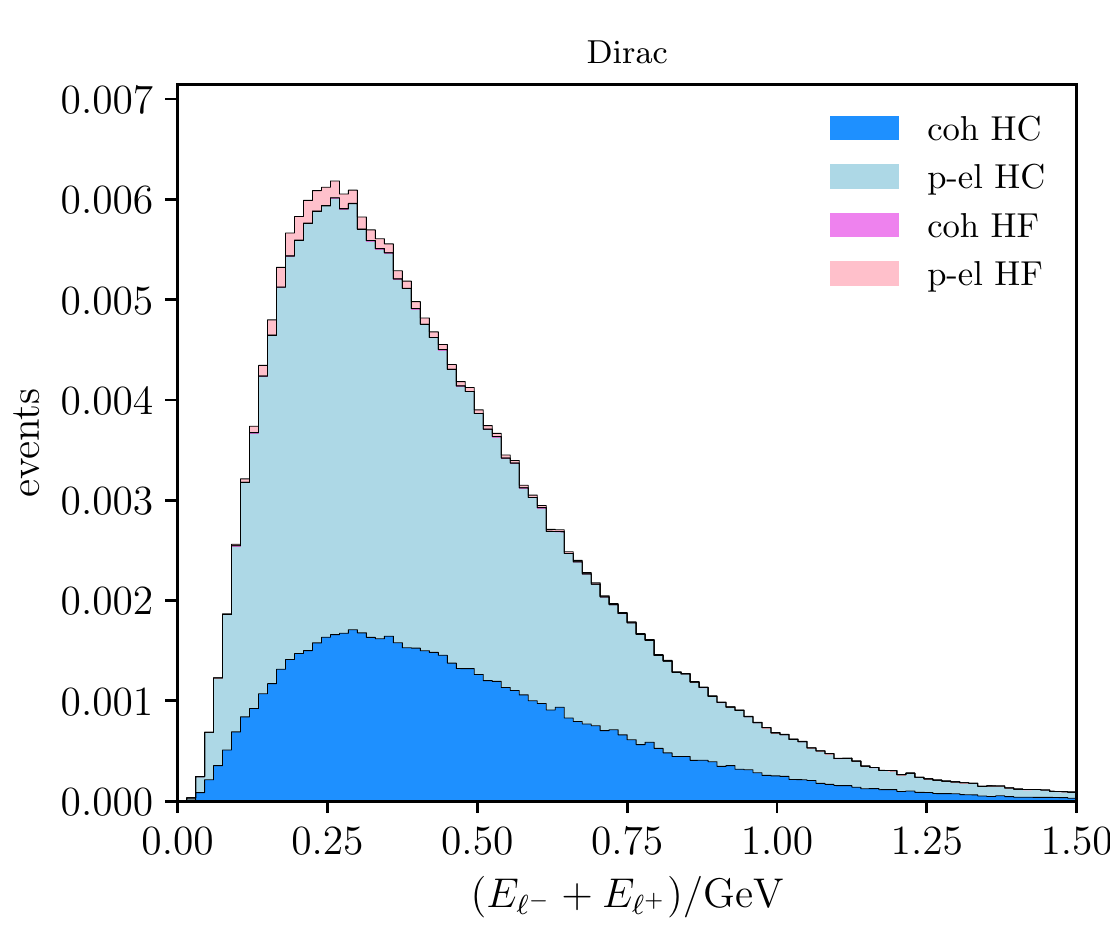}
    \includegraphics[width=0.32\textwidth]{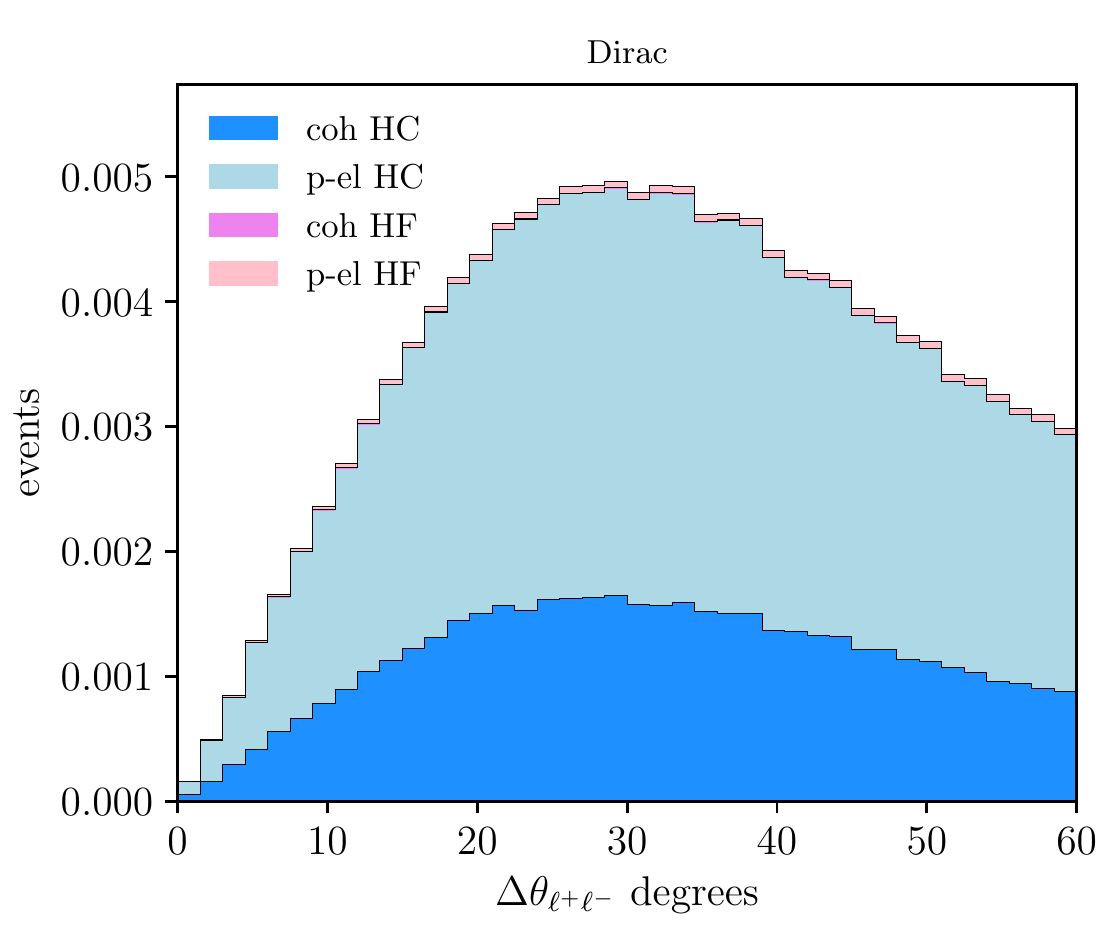}
    \includegraphics[width=0.32\textwidth]{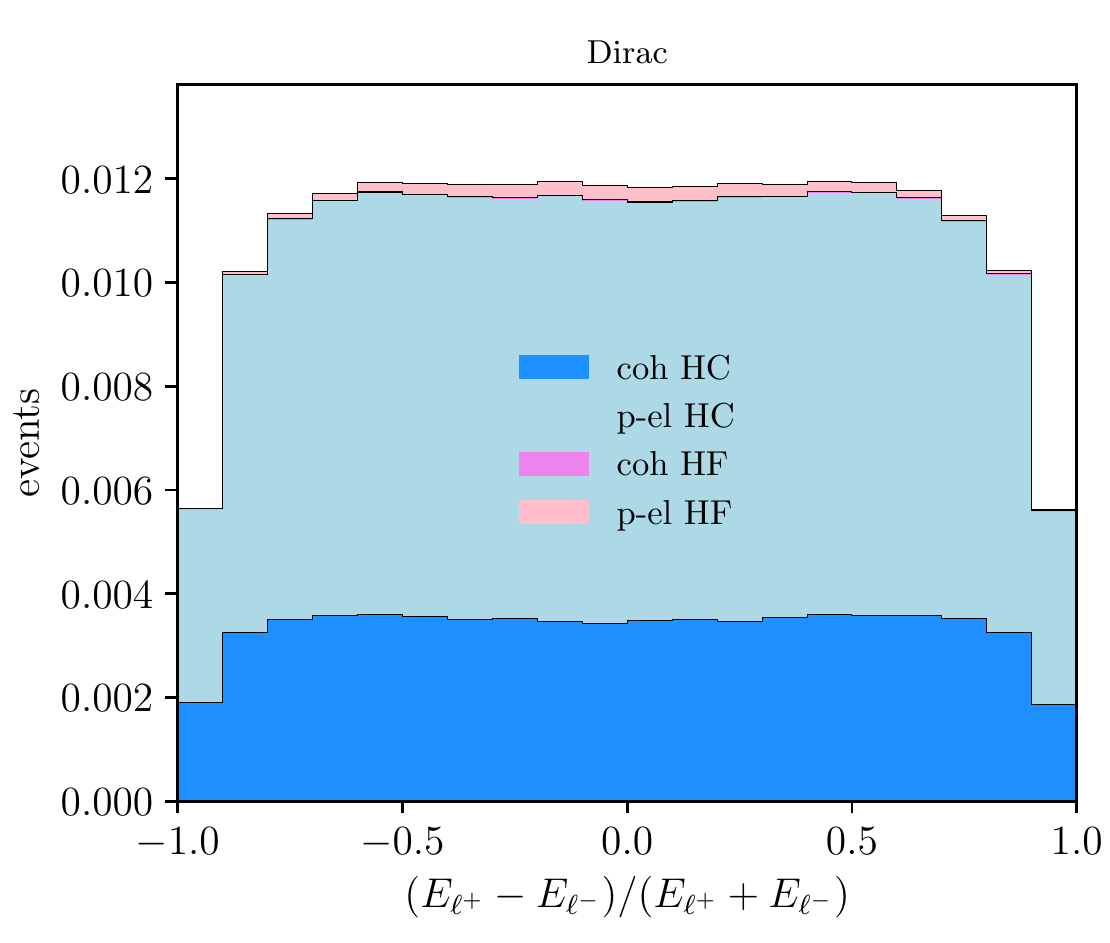}
    \\
    \includegraphics[width=0.32\textwidth]{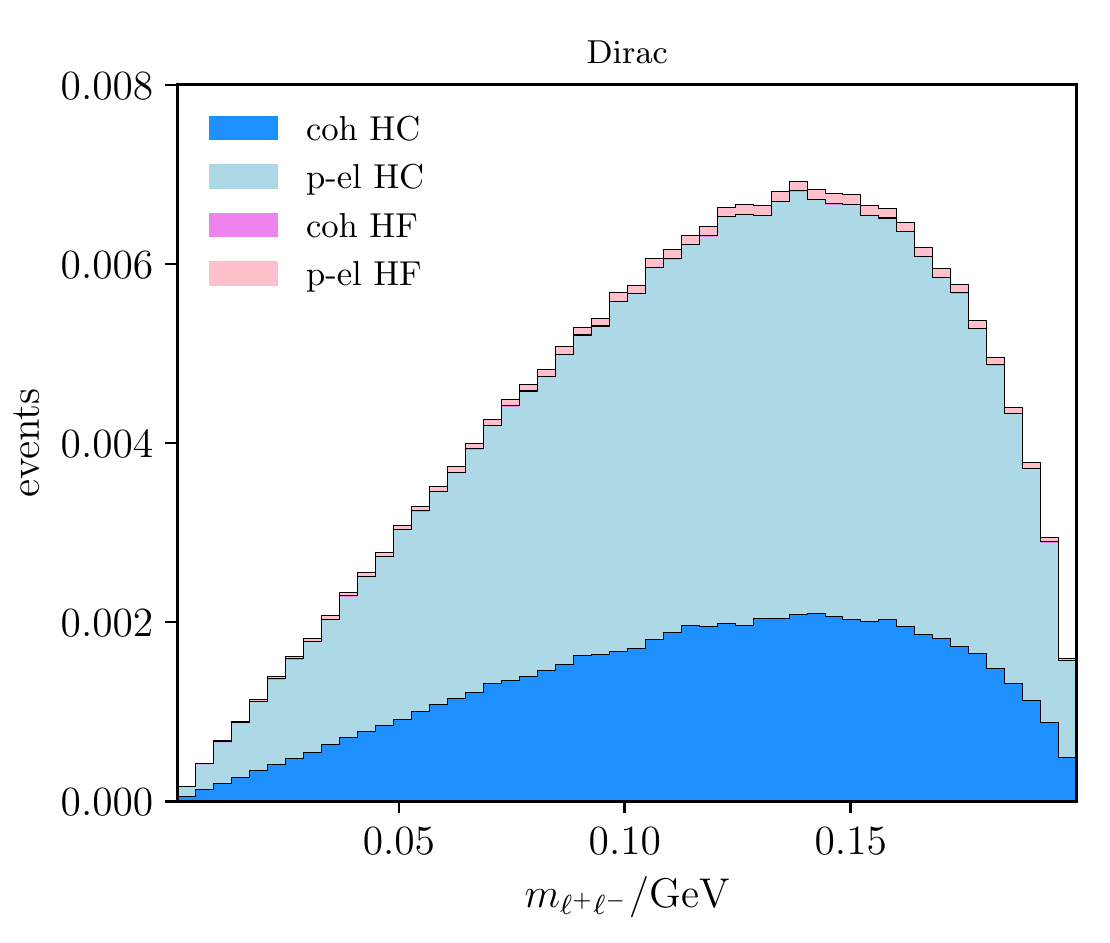}
    \includegraphics[width=0.32\textwidth]{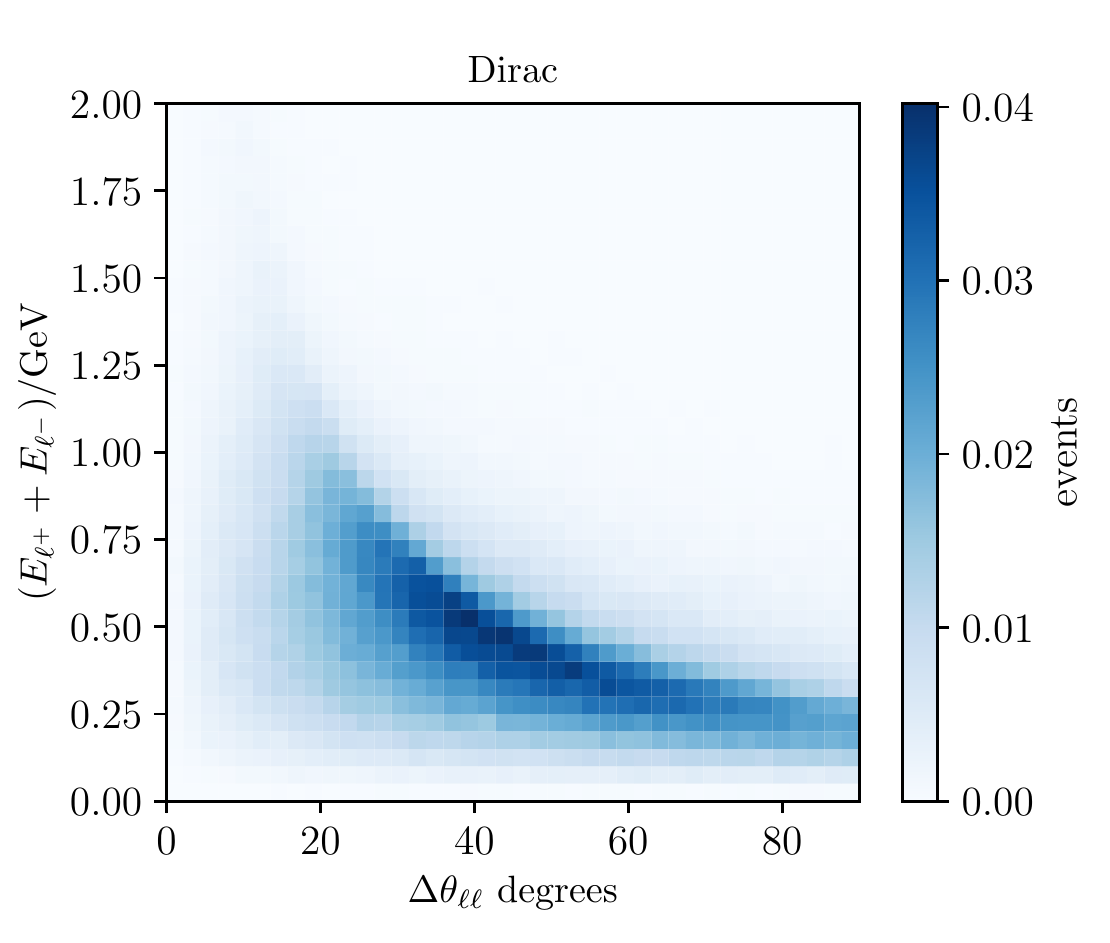}
    \includegraphics[width=0.32\textwidth]{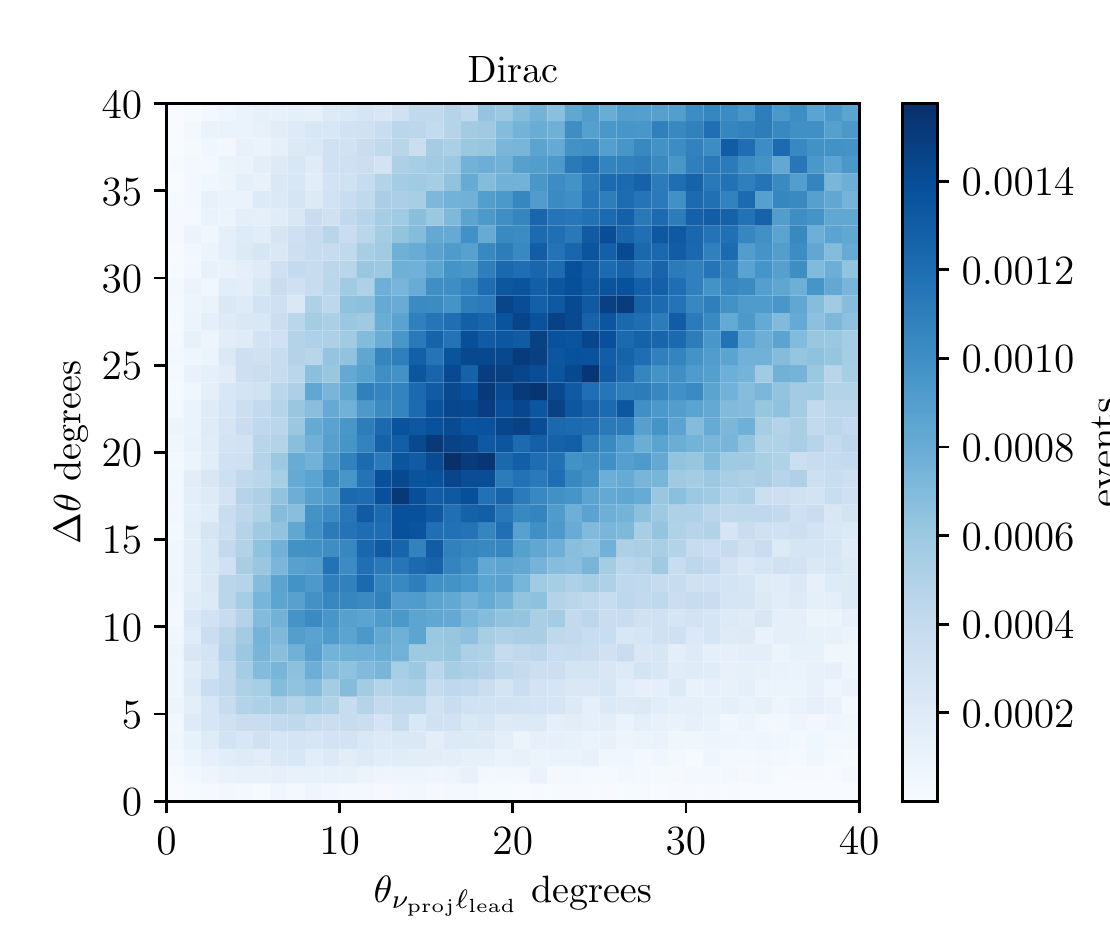}
    \caption{Stacked histograms of the kinematical distributions of the final states produced in the neutrino-nucleus upscattering. Heavy neutrino is a Dirac particle with model parameters taken to be: $g_D = 1$, $\epsilon = 10^{-4}$, $m_5 = 300$~MeV,  $m_4 = 100$~MeV, and  $m_{Z^\prime} = 1.25$~GeV. See  text for definitions.
    \label{fig:distributions_zprime_dir_heavy}}
\end{figure*}

For the sake of illustration, we will assume that neutrino upscattering is mediated only by the $Z$ boson and the dark photon.
For a massless projectile incident on a nucleon of energy $E_p$, we can write the single differential cross section as
\begin{widetext}
\begin{align}\label{eq:sigma_nuN}
    \frac{\dd \sigma}{\dd Q^2} (\nu_i \mathcal{H}  \to N_j^h \mathcal{H}) =
    \frac{1}{64 \pi s |\vec{p}_\nu^{\, \, \rm \small CM}|^2} 
   {\rm L}_{\mu\nu}^h \Bigg[& 
        \frac{(d_{ij} d_{V}^{\mathcal{H}})^2 }{(Q^2 + M_{Z^\prime}^2)^2}{\rm H}^{\mu\nu}_{\rm EM} 
    +
    \frac{(c_{ij} c_{V}^{\mathcal{H}})^2 }{(Q^2 + M_{Z}^2)^2}{\rm H}^{\mu\nu}_{\rm NC}
    +
    \frac{ 2 c_{ij} d_{ij} c_{V}^{\mathcal{H}} d_{V}^{\mathcal{H}}}{(Q^2 + M_{Z}^2)(Q^2 + M_{Z^\prime}^2)}{\rm H}^{\mu\nu}_{\rm int} \Bigg],
\end{align}
\end{widetext} 
where the leptonic tensor is given by
\begin{align}
{\rm L}_{\mu\nu}^h &= -\frac{g^{\mu\nu}}{2} \left[ m_\nu^2 + m_N^2 + Q^2 - 2 h (k\cdot s)  m_N \right] 
\\\nonumber
&\qquad - (k^\mu s^\nu + k^\nu s^\mu) h \,m_N 
+ (k^\mu k^{\prime\nu} + k^\nu k^{\prime\mu}) 
\\\nonumber
&\qquad + i \epsilon^{\mu \nu \rho\sigma} \left[ k^\rho k^{\prime\sigma} - h m_N k^\rho s^{\prime\sigma} \right]  ,
\end{align}
which in the limit of $m_\nu = m_N = 0$ is the usual weak leptonic tensor ${\rm L}_{\mu\nu} = -(Q^2/2) g^{\mu\nu} + (k^\mu k^{\prime\nu} + k^\nu k^{\prime\mu}) + i \epsilon^{\mu \nu \rho\sigma} k^\rho k^{\prime\sigma}$. 
The hadronic tensors are determined for each scattering regime assuming a spin-1/2 nucleus and nucleon.

\paragraph{Form factors}
For coherent scattering, we use only the electric form factor for $\gamma$, $Z$, and $Z^\prime$ exchange (with the appropriate interaction vertex), while for a scalar exchange we implement the formalism in \refref{Cline:2013gha}.
For the proton and neutron elastic cases, we implement the relevant Dirac, Pauli, and Axial form factors.
Nuclear form factors are given by either the Fourier-Bessel parametrization~\cite{Fricke:1995zz,DeVries:1987atn,DeJager:1974liz} when nuclear data is available (given in machine-friendly format by~\refref{VT_NDT}), or by a Fermi-Symmetrized function when they are not. 
These are described in more detail in the Appendix of \refref{Kamp:2022bpt}.

\paragraph{Helicity}
As we assume the neutrino beam is fully polarized, the only helicity that we keep track of is that of the produced HNL. The scattering can then be helicity flipping, where the outgoing HNL has the opposite helicity of the initial projectile, or helicity-conserving, where they are the same. Depending on the interaction, one of these channels dominates over the other. For instance, scalar interactions and transition magnetic moments couple two separate chiralities and, therefore, lead to dominance in the helicity flipping channels. The opposite is true for vectorial interactions mediated by the dark photon.  The decay of the HNL is then performed using the relevant helicity state produced. This allows for a proper comparison of the kinematics in the case of Majorana and Dirac neutrinos.

\paragraph{Additional contributions}
If one, or two, of the heavy neutrinos are long-lived with respect to the baseline of the experiment, their flux may also be non-negligible at the detector.
In addition to their decays-in-flight (which are not included in \darknus), they can also initiate upscattering processes. 
It is worthwhile to estimate such processes since the upscattering cross section involving only heavy neutrinos is no longer proportional to portal parameters but rather to the large couplings between $N_i$ and $N_j$ pairs.
While \darknus does not consider such processes, the user can obtain an estimate of this rate by reweighing light neutrino upscattering.

For instance, in models with $N_4$ and $N_5$ and a dark photon, the $N_4$-initiated process $N_4 \mathcal{H} \to N_5 \mathcal{H} \to \nu e^+e^- \mathcal{H}$ can be estimated by re-weighting the process $\hat{\nu}_\mu \mathcal{H} \to N_5 \mathcal{H}$.
Taking into account that the flux of $N_4$ would depend on its parent meson and including only one polarization structure, the ratio of $N_4$- and $\nu_\mu$-initiated upscattering events is
\begin{equation}
R = \sum_{M} \left(\frac{d_{45}}{d_{\mu 5}}\right)^2 \rho\left(\frac{m_4^2}{m_M^2},\frac{m_\mu^2}{m_M^2}\right) \frac{\Phi_{M}(E_\nu)}{\Phi(E_\nu)} P_{\rm surv},    
\end{equation}
where $\rho(a, b) = |U_{\mu 4}|^2 (a+b-(a-b)^2)\sqrt{\lambda(1,a,b)}$, $\lambda(1,a,b)$ is the K\"all\'en function, $\Phi_{M}(E_\nu)$ is the flux of neutrinos from the meson $M$, $\Phi(E_\nu)$ is the total neutrino flux used in the simulation, and $ P_{\rm surv}$ is the survival prolbability of $N_4$.
Note that this ratio scales as $|U_{\mu 4}|^2/|U_{\mu 5}|^2$ and so this additional contribution is not fully determined by the parameters that enter into the $\nu_\mu$ upscattering included in \darknus.
In addition, in many cases $P_{\rm surv}$ is negligible, or $|U_{\mu 4}|^2$ is constrained to be very small by decay-in-flight searches. 

\subsection{Event rate}

The differential experimental event rate for dileptons or single photons produced in the subsequent decay of the HNL is given by
\begin{equation}\label{eq:dNdQ2dPS}
    \frac{\dd N(E_\nu)}{\dd Q^2\,{\rm dPS}} = \frac{\dd\sigma(E_\nu)}{\dd Q^2} \left(\frac{\dd \Gamma_{N \to \nu X}}{{\rm dPS}}\frac{1}{\Gamma_{\rm tot}}\right),
\end{equation}
where $\dd \Gamma_{N\to \nu X}/{\rm dPS}$ is the differential decay rate of the HNL into a neutral lepton and $X=e^+e^-, \mu^+\mu^-$ or $\gamma$, and $\Gamma_{\rm tot}$ is the total decay rate of the HNL. 
The whole quantity in parentheses can be thought of as a differential branching ratio.

In the final dataframe, \darknus stores the left-hand side of \Cref{eq:dNdQ2dPS} as \texttt{"w\_event\_rate"}, the differential cross section as \texttt{"w\_flux\_avg\_xsec"}, and the differential decay rate $\dd \Gamma_{N\to \nu X}/{\rm dPS}$ as \texttt{"w\_decay\_rate\_i"}, where $i=0$ or $1$, depending on the process. 
The event rate is properly normalized to the experimental exposure and number of scattering targets, such that the total event rate can be obtained by  \mintinline{python}|df.w_event_rate.sum()|. The total decay rate of the HNLs is stored in the attributes of the dataframe.
\begin{figure*}[t]
    \centering
    \includegraphics[width=0.32\textwidth]{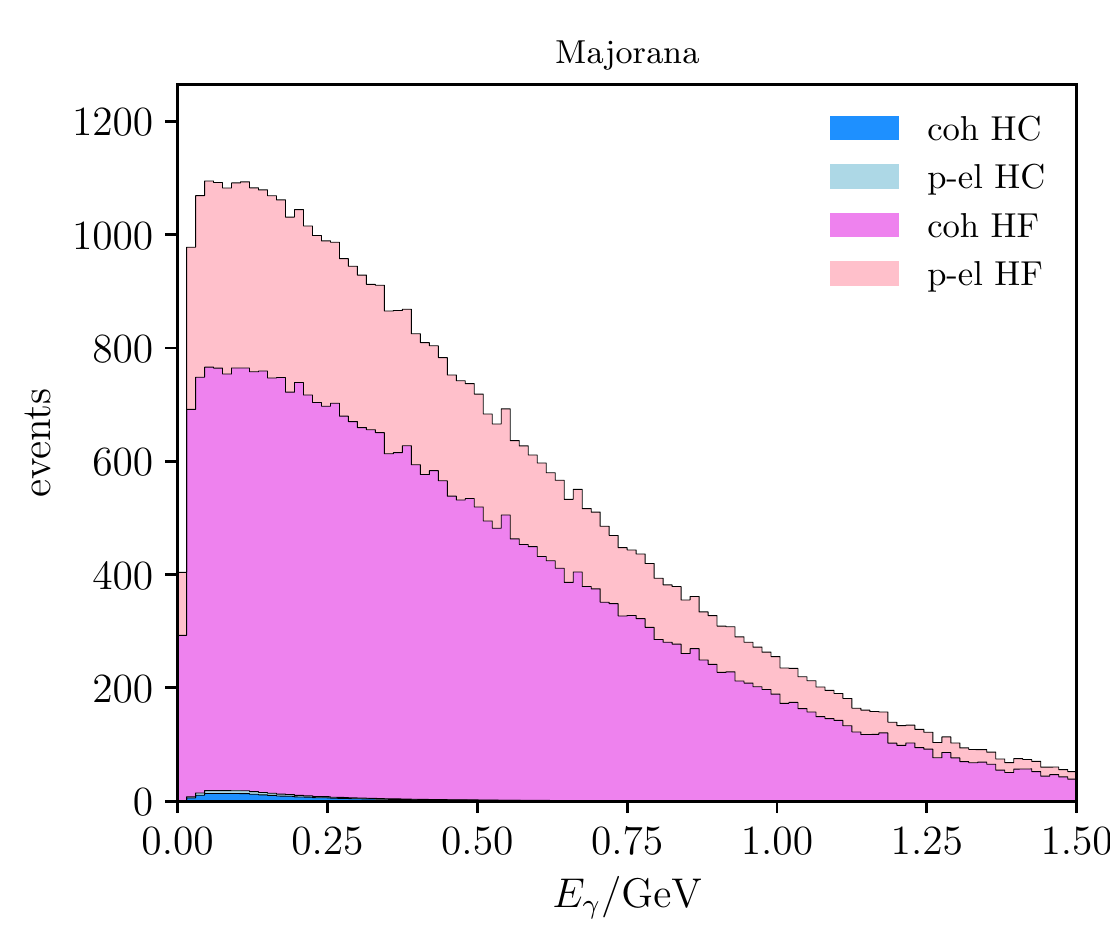}
    \includegraphics[width=0.32\textwidth]{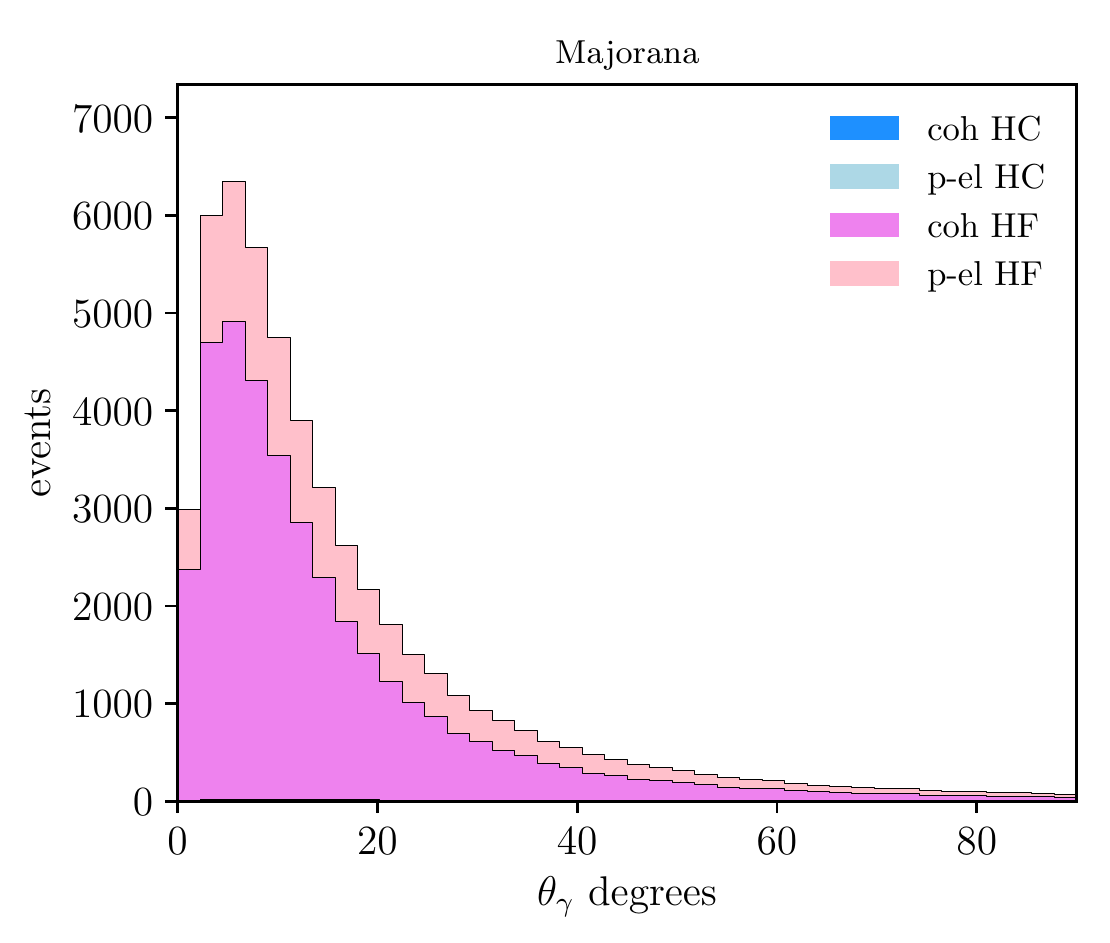}
    \includegraphics[width=0.32\textwidth]{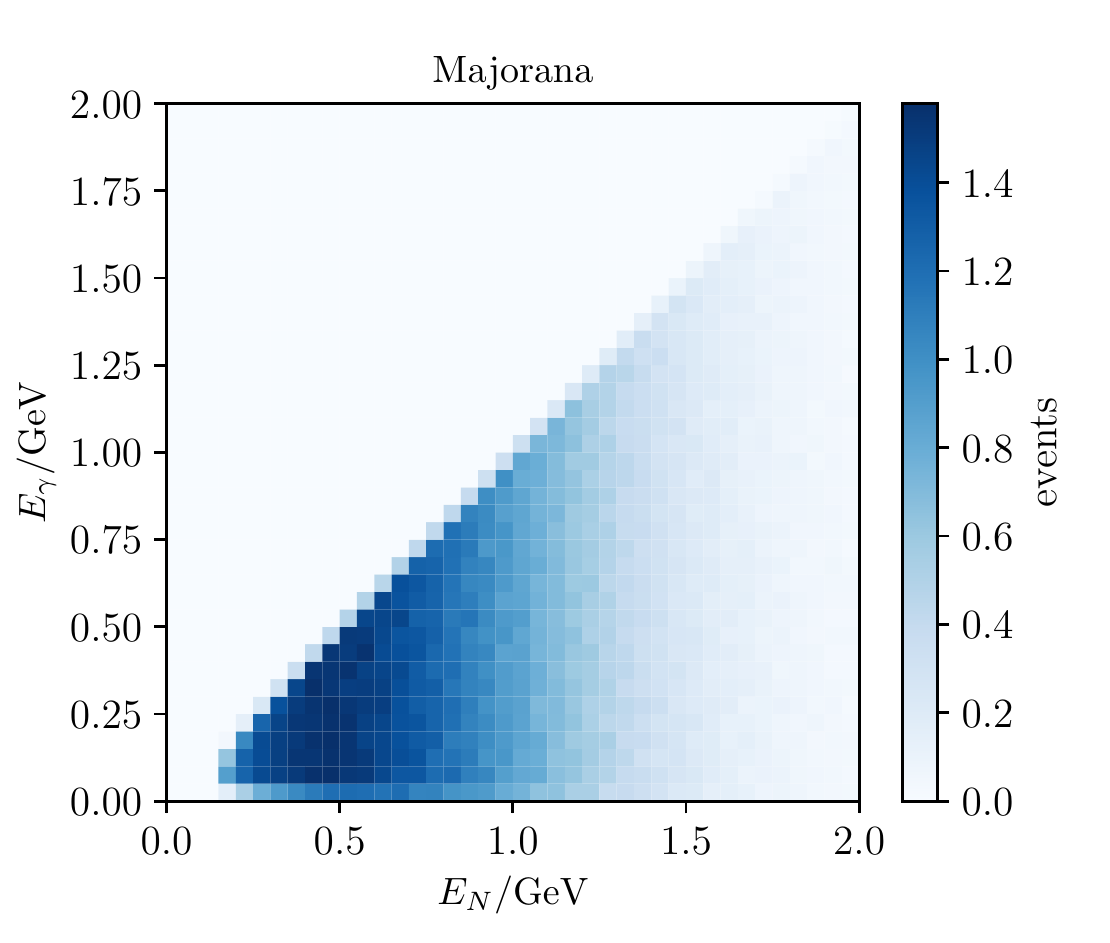}
    \\
    \includegraphics[width=0.32\textwidth]{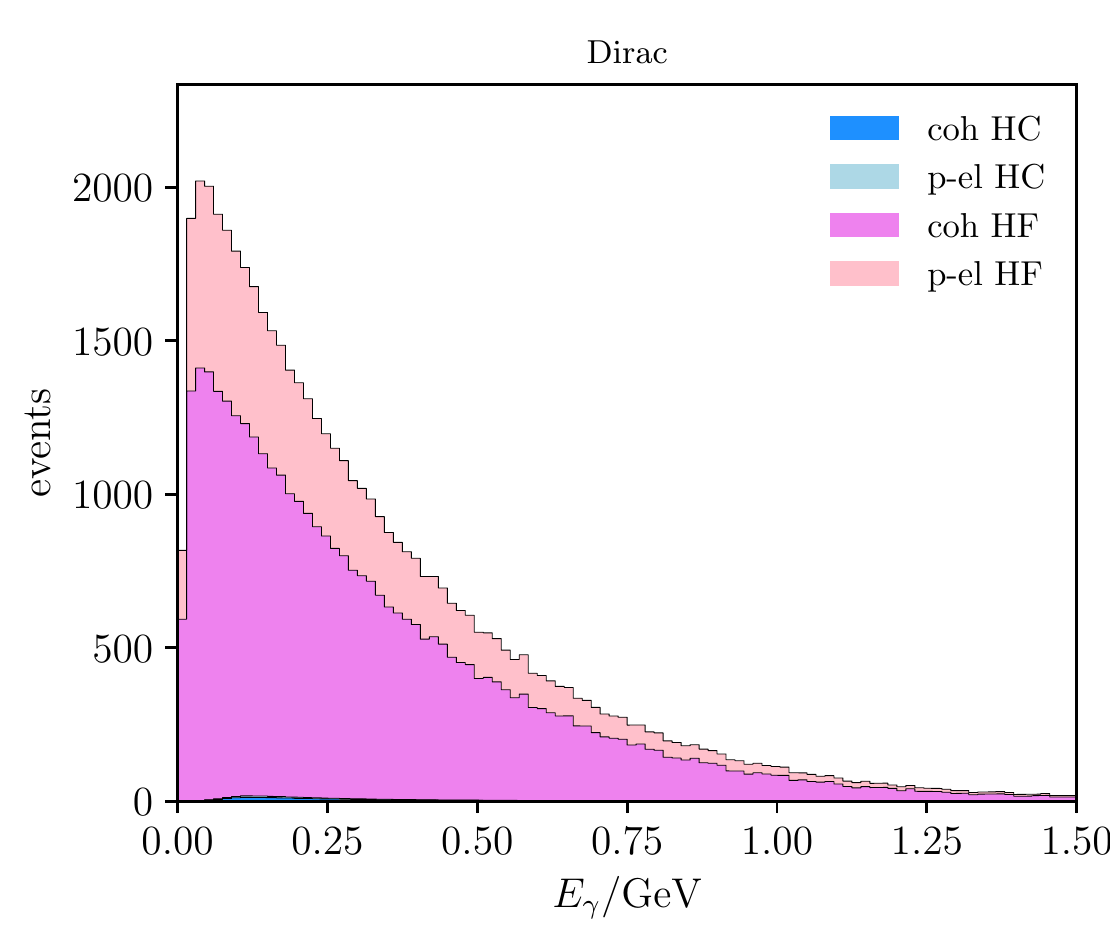}
    \includegraphics[width=0.32\textwidth]{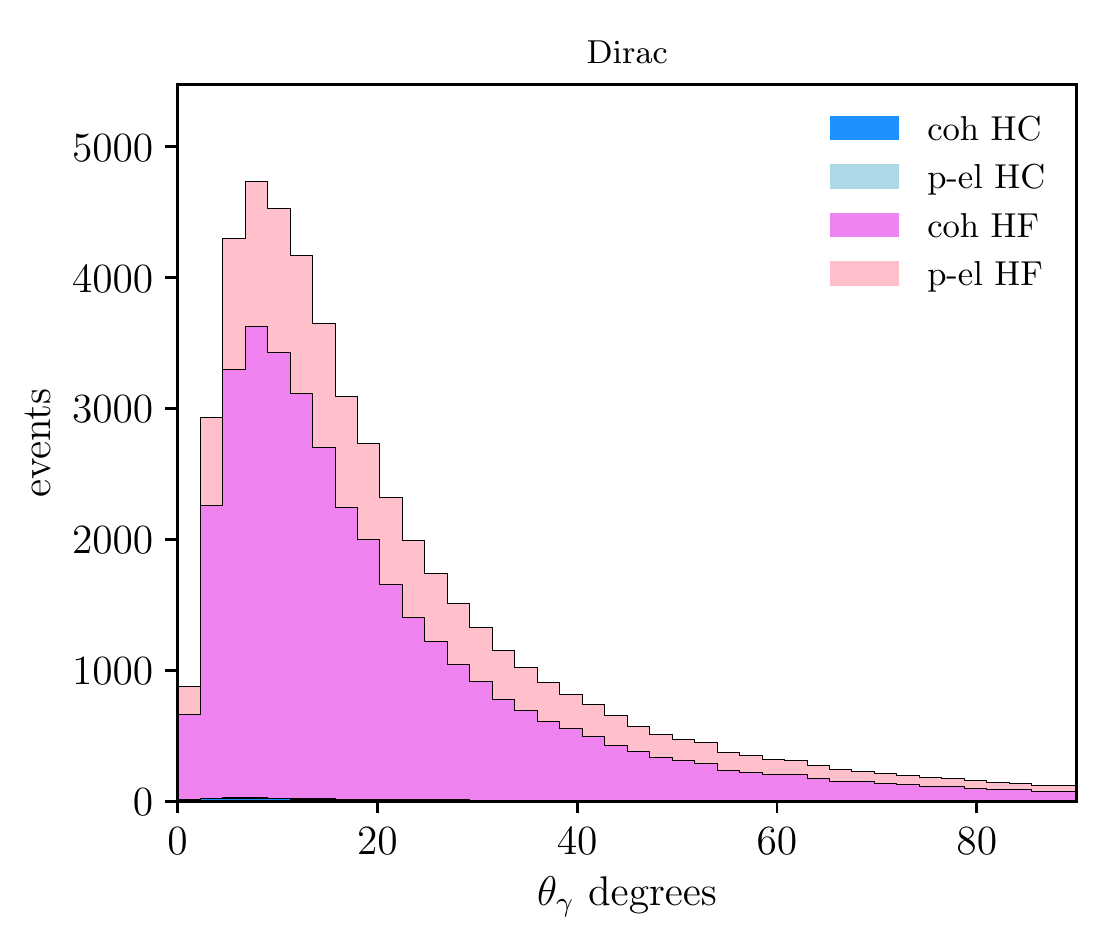}
    \includegraphics[width=0.32\textwidth]{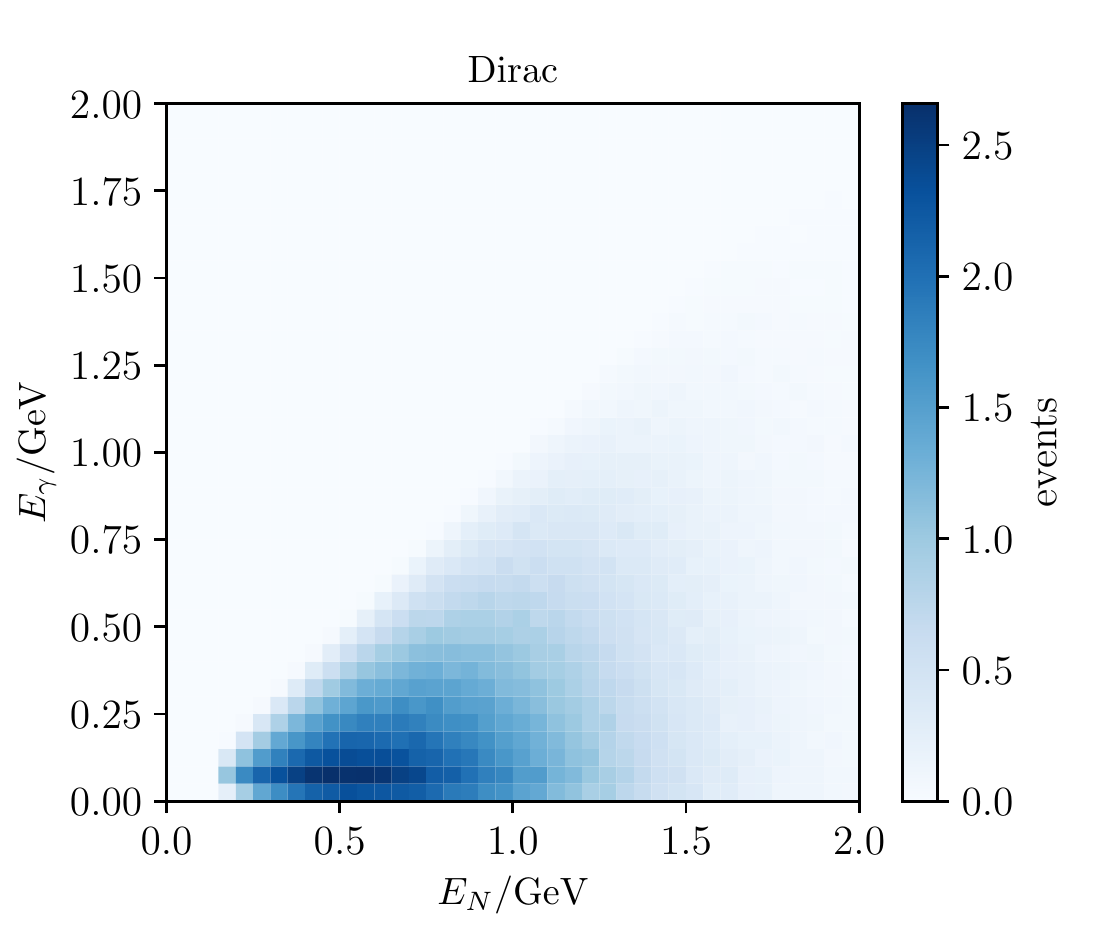}
    \caption{Stacked histograms with the kinematical distributions of the final states produced in neutrino-nucleus upscattering for a transition magnetic moment model. For this point, we use a Majorana heavy neutrino for the top row and a Dirac heavy neutrino for the bottom row with $m_4=150$~MeV and $\mu^{\rm tr}_{\mu 4} = 1$~PeV$^{-1}$.
    See the text for definitions.
    \label{fig:distributions_mutr}}
\end{figure*}


\section{Examples}\label{sec:examples}

In this section, we show some simple examples of the usage of \darknus. 
First, we note that earlier versions of this code have been used in the following studies: Refs.~\cite{Ballett:2019pyw,Abdullahi:2020nyr} with MiniBooNE predictions, Ref.~\cite{Arguelles:2018mtc} studying MINERvA, and Ref.~\cite{Arguelles:2021dqn} studying ND280, the near detector of T2K.
For this version, we have provided comprehensive examples in the \texttt{Examples/} folder of the repository in the form of documented Jupyter notebooks.

We now show how to compute cross sections and generate events, showing some kinematical distributions.

\subsection{Cross sections}

It is possible to get differential and total cross sections from \darknus. 
Differential cross sections are always given in terms of the momentum exchange $Q^2$, with limits given by \mintinline{python}|upscattering_Q2min(Enu, mHNL, M)| and \mintinline{python}|upscattering_Q2max(Enu, mHNL, M)|, where Enu, mHNL, and M stand for the neutrino energy, heavy neutrino mass, and nuclear/nucleon mass, respectively.

\vspace{5ex}
\begin{pythoncode}
import numpy as np
import pandas as pd

import DarkNews as dn
from DarkNews import GenLauncher
from DarkNews import phase_space
from DarkNews import pdg

# targets
C12    = dn.detector.NuclearTarget("C12")

# model
SM_like = dn.model.ThreePortalModel(m4 = 0.001, Umu4=1e-3, epsilon=0, gD=0)

# xsec calculators
calc_NC_coh   = dn.UpscatteringProcess(
TheoryModel = SM_like, 
scattering_regime = 'coherent', 
nu_projectile=pdg.numu, 
nu_upscattered=pdg.neutrino4, 
helicity='conserving', 
nuclear_target=C12) 

# calculating it
xsec_NC = calc_NC_coh.total_xsec(Enu = np.geomspace(0.01,2e2,50), NEVAL=1000, NINT=10)
\end{pythoncode}

\subsection{Upscattering at MiniBooNE}

We now generate events for the MiniBooNE experiment in neutrino mode for a $3+2$ heavy neutrino model. 
The resulting kinematic distributions for a model with a light dark photon mediator can be found in \Cref{fig:distributions_zprime_maj} and \Cref{fig:distributions_zprime_dir} for Majorana and Dirac HNLs, respectively, 
and in \Cref{fig:distributions_mutr} for a transition magnetic moment model with Dirac neutrinos.
In all cases, the plots are easily obtained with the \mintinline{python}|--make-summary-plots| argument. 

For the light $Z^\prime$ cases in \Cref{fig:distributions_zprime_dir} and \Cref{fig:distributions_zprime_maj} we show the sum of the $e^+e^-$ energy, the angle between the projectile neutrino (in the $\hat{z}$-direction) and the sum of the $e^+e^-$ momenta referred to as $\theta_{(\ell\ell)\nu_{\rm proj}}$, the opening angle between the two charged lepton $\theta_{\ell\ell}$, the angle between the projectile neutrino and the leading (highest energy) charged lepton $\theta_{\nu_{\rm proj} \ell}$, and the energy of the parent HNL $E_N$. 
For the heavy $Z^\prime$ in \Cref{fig:distributions_zprime_dir_heavy}, we also show the energy asymmetry $(E_{\ell^+}-E_{\ell^-})/(E_{\ell^+}+E_{\ell^-})$ and the invariant mass $m_{\ell^+\ell^-}$.
The Dirac-Majorana differences in the heavy $Z^\prime$ case are much less pronounced and appear mostly in the energy asymmetry -- for Dirac, the center region is slightly concave, while for Majorana it is slightly convex.
For the invariant mass, it leans backward for Majorana, and forward for Dirac.
The total energy and opening angles remain very similar.

For the transition magnetic moment case in \Cref{fig:distributions_mutr}, we show the photon energy $E_\gamma$, the angle of the photon with respect to the projectile neutrino $\theta_\gamma$, and the energy of the parent heavy neutrino $E_N$.
The Dirac-Majorana differences in this case are also important for the total energy and angle.

In \Cref{fig:distributions_zprime_maj} and \Cref{fig:distributions_zprime_dir}, we can see that the dominant mode of upscattering is the helicity-conserving one due to the vector nature of the dark photon interactions with matter.
The dilepton angles with respect to the neutrino beam are also rather forward, as expected.
Dirac HNLs lead to less forward angles and lower energy signatures, as well as other subtle effects.
In \Cref{fig:distributions_mutr}, on the other hand, the dominant upscattering mode is the helicity-flipping one since the transition magnetic moment interaction is of a similar chiral structure to the scalar case.
The photon energy is lower energy for the same reason: the photon is preferentially emitted in the backward direction and tends to be lower energy and less forward.

\section{Conclusions}\label{sec:conclusions}

We have presented \darknus, an easy-to-install, easy-to-use, \python-based Monte-Carlo generator for new physics in neutrino-nucleus scattering.
The code models neutrino upscattering to heavy neutrinos and their subsequent decays into dilepton pairs or single photons at neutrino experiments.
New physics models may be defined by the user and can contain up to three heavy neutrinos. 
Additional interactions may be defined for the heavy neutrinos including e.g. a dark photon, a dark scalar particle, or a transition magnetic moment.
We model coherent nuclear scattering, as well as free proton- and neutron-elastic scattering (including bound nucleons without nuclear modifications).
The generator is able to output events to several formats, including the internally-used \pandas dataframe and \numpy ndarrays.
Events may also be saved to HEPevt and HepMC2/3 files, both of which are frequently used by experimental collaborations.

We have provided examples of how to obtain differential and total cross sections, as well as kinematical distributions of $e^+e^-$, $\mu^+\mu^-$, and single-photon events.

In recent years, interest in the development of new event generators for neutrino experiments has grown and a number of sophisticated new tools join the existing event generators. While the implementation of new physics models in these newer tools exceeds the traditional generators in ease, there is a need for lighter-weight generators to be used by both the phenomenology and experimental communities. The power of \darknus lies in its simplicity and speed and we believe it will be a valuable tool in the search for new physics in current and future neutrino experiments.

\acknowledgments
We are grateful to G.~Peter~Lepage for correspondence and for implementing new features in \href{https://vegas.readthedocs.io/en/latest/tutorial.html#vegas-jacobian}{{\vegas}}.
We would like to thank Nic\`olo~Foppiani, Joshua~Isaacson, Georgia Karagiorgi, and Mark~Ross-Lonergan for suggestions that helped improve the code.
The research of M.H. was supported in part by Perimeter Institute for Theoretical Physics. Research at Perimeter Institute is supported by the Government of Canada through the Department of Innovation, Science and Economic Development and by the Province of Ontario through the Ministry of Research, Innovation and Science. 
The research of J.H.Z. has received funding / support from the European Union’s Horizon 2020 research and innovation programme under the Marie Skłodowska-Curie grant agreement No 860881-HIDDeN.
The research of A.A is supported by Fermi Research Alliance, LLC under Contract No. DE-AC02-07CH11359 with the U.S. Department of Energy, Office of Science, Office of High Energy Physics.

\appendix

\section{Formats, input, and outputs}
\label{sec:appendix}
The input parameters for the generator are shown in \Cref{tab:input_parameters}, while the input parameters that define the underlying new physics model to be used are shown in \Cref{tab:model_inputs} for the generic model of \Cref{eq:Ninteractions} and \Cref{eq:SMinteractions}, as well as for the three portal model of \cite{Abdullahi:2020nyr}.
The experiment definition parameters are listed in \Cref{tab:input_parameters,tab:exp_definitions}. 

The format of the output \pandas dataframe is shown in \Cref{tab:pandas_format}. 
This dataframe is the main result of the generator and is used internally in \darknus.
\clearpage
\LTcapwidth=\textwidth
\renewcommand{\arraystretch}{1.3}
\begin{longtable*}[t]{|>{\tt}l|>{\em}c|c|>{\arraybackslash}p{10.4 cm}|}
    \hline
        \normalfont\textbf{Parameter} & \normalfont\textbf{Type} & \normalfont\textbf{Default} & \normalfont\textbf{Description} \\
    \hline
    \hline
        \multicolumn{4}{|c|}{\textbf{Experiment}} \\
    \hline
        exp & string & {\normalfont\mintinline{python}|"miniboone_fhc"|} & The experiment to consider: see text for more information \\
    \hline
    \hline
        \multicolumn{4}{|c|}{\textbf{Monte-Carlo scope}} \\
    \hline
        nopelastic & bool & {\normalfont\mintinline{python}|False|} & Do not generate proton elastic events     \\ 
        nocoh & bool & {\normalfont\mintinline{python}|False|} & Do not generate coherent events           \\ 
        noHC & bool & {\normalfont\mintinline{python}|False|} & Do not include helicity conserving events \\ 
        noHF & bool & {\normalfont\mintinline{python}|False|} & Do not include helicity flipping events   \\
        decay\_products & string & {\normalfont\mintinline{python}|"e+e-"|} & Decay process of interest, choose among \mintinline{python}|["e+e-","mu+mu-","photon"]| \\
        nu\_flavors & list of string & {\normalfont\mintinline{python}|"nu_mu"|} & Neutrino projectile, a list with elements among \mintinline{python}|["nu_e","nu_mu","nu_tau","nu_e_bar","nu_mu_bar","nu_tau_bar"]| \\
        sample\_geometry & bool & {\normalfont\mintinline{python}|True|} & Whether to sample the detector geometry using DarkNews. If False or a geometry function is not found, the upscattering position is assumed to be $(0,0,0,0)$. \\
        enforce\_prompt & bool & {\normalfont\mintinline{python}|False|} & Force particles to decay promptly \\
    \hline
    \hline
        \multicolumn{4}{|c|}{\textbf{\vegas integration options}} \\
    \hline
        neval & int & 10000 & Number of evaluations of integrand           \\ 
        nint & int & 20 & Number of adaptive iterations                \\ 
        neval\_warmup & int & 1000 & Number of evaluations of integrand in warmup \\ 
        nint\_warmup & int & 10 & Number of adaptive iterations in warmup      \\
        seed & int & \mintinline{python}|None| & \numpy random number generator seed used in \vegas       \\
    \hline
    \hline
        \multicolumn{4}{|c|}{\textbf{Output format options}} \\
    \hline
        pandas & bool & {\normalfont\mintinline{python}|True|} & Save \mintinline{python}|pandas.DataFrame| as \texttt{.pckl} file                                \\ 
        parquet & bool & {\normalfont\mintinline{python}|False|} & Save \mintinline{python}|pandas.DataFrame| as \texttt{.parquet} file (\mintinline{python}|engine=pyarrow|)                                \\ 
        numpy & bool & {\normalfont\mintinline{python}|False|} & Save events in \texttt{.npy} files                                            \\ 
        hepevt & bool & {\normalfont\mintinline{python}|False|} & Save events in HEPEVT-formatted text files                             \\ 
        hep\_unweight & bool & {\normalfont\mintinline{python}|False|} & Unweight events when printing in any of the HEP formats (needs large statistics) \\ 
        unweighted\_hep\_events & int & 100 & Unweight events when printing in any of the HEP formats (needs large statistics) \\ 
        sparse & bool & {\normalfont\mintinline{python}|False|} & if True, save only the neutrino energy, charged lepton or photon momenta, and weights. Not supported for HEP formats.    \\
        make\_summary\_plots & bool & {\normalfont\mintinline{python}|False|} & if True, generates summary plots of kinematics in the `path` \\
        path & string & {\normalfont\mintinline{python}|"./"|} & Path where to save run's outputs\\ 
\hline
    \hline
        \multicolumn{4}{|c|}{\textbf{Output verbosity}} \\
    \hline
        loglevel & string & {\normalfont\mintinline{python}|"INFO"|} & Logging level, choose among \mintinline{python}|["INFO", "WARNING", "ERROR", "DEBUG"]| \\
        verbose & bool & \mintinline{python}|False| & Verbose for logging                          \\ 
        logfile & string & {\normalfont\mintinline{python}|None|} & Path to log file; if not set, use std output \\
    \hline
\caption{The input parameters that can be set in \darknus. We separate them into broad categories regarding the behavior of the generator.\label{tab:input_parameters}}
\end{longtable*}

\clearpage
\renewcommand{\arraystretch}{1.3}
\begin{table*}[t]
\begin{tabular}{|>{\tt}l|>{\arraybackslash}p{14.75 cm}|}
    \hline
        \normalfont\textbf{Parameter} & \normalfont\textbf{Description} \\
    \hline
    \hline
        \multicolumn{2}{|c|}{\textbf{Dark sector spectrum}} \\
    \hline
        m\{i\} & Mass of the heavy neutrinos ($i=4,5,6$, default $m_4 = 150$~MeV) \\ 
        mzprime &  Mass of dark photon $Z^\prime$ (default $m_{Z^\prime}=1.25$) \\ 
        mhprime & Mass of dark scalar $h^\prime$ \\ 
        HNLtype & Dirac or Majorana HNLs? (default "dirac")\\ 
    \hline
    \hline
        \multicolumn{2}{|c|}{\textbf{Common model-independent parameters}} \\
    \hline
        mu\_tr\_\{ij\} & Transition magnetic moment defined in \Cref{eq:Ninteractions} ($i = e,\mu,\tau,4,5,6$)   \\ 
        s\_\{ij\} & Scalar couplings defined in \Cref{eq:Ninteractions} ($i = e,\mu,\tau,4,5,6$)   \\ 
    \hline
    \hline
        \multicolumn{2}{|c|}{\textbf{Generic model parameter setting}} \\
    \hline
        c\_\{ij\} & Neutral lepton couplings of the SM Z boson ($i = e,\mu,\tau,4,5,6$)   \\ 
        d\_\{ij\} & Neutral lepton couplings of the dark photon $Z^\prime$ ($i = e,\mu,\tau,4,5,6$)   \\ 
    \hline
        c\{f\}V & SM fermions vector couplings of the SM Z boson ($f = e, u, d$)  \footnotemark[3]   \\ 
        c\{f\}A & SM fermions axial-Vector couplings of the SM Z boson ($f = e, u, d$)  \footnotemark[3]   \\ 
        d\{f\}V & SM fermions vector couplings of the dark photon $Z^\prime$ ($f = e, u, d$)  \footnotemark[3]   \\
        d\{f\}A & SM fermions axial-vector couplings of the dark photon $Z^\prime$ ($f = e, u, d$)  \footnotemark[3]   \\
        d\{f\}S & SM fermions scalar couplings of the dark scalar $h^\prime$ ($f = e, u, d$)  \footnotemark[3]   \\
        d\{f\}P & SM fermions pseudo-scalar couplings of the dark scalar $h^\prime$ ($f = e, u, d$)  \footnotemark[3]   \\
    \hline
    \hline
        \multicolumn{2}{|c|}{\textbf{Three-portal model parameter setting}} \\
    \hline
        ue\{i\} & Mixing element between electron flavor and heavy neutrinos \\ 
        umu\{i\} &  Mixing element between muon flavor and heavy neutrinos ($|U_{\mu 4}| = 10^{-3}$)\\ 
        utau\{i\} & Mixing element between tau flavor and heavy neutrinos  \\ 
        ud\{i\} & Mixing element between dark flavor and heavy neutrinos ($|U_{D 4}| = 1$)\\
    \hline
        gD & $\mathrm{U}(1)_\textup{d}$ dark coupling ($g_\textup{D} = 1$ ) \\ 
        alphaD & $\alpha_\textup{D} = g_\textup{D}^2 \, / \, (4\pi)$ \footnotemark[1]\\
        epsilon & Kinetic mixing parameter ($\varepsilon = 10^{-3}$) \\ 
        epsilon2 & $\varepsilon^2$   \footnotemark[2] \\ 
        alpha\_epsilon2 & $\alpha \varepsilon^2$ \footnotemark[2] \\ 
        chi & $\chi = \varepsilon \, / \, c_w$ \footnotemark[2] \\
        theta & Scalar mixing $\theta$ between the Higgs and $h^\prime$ \\
    \hline    
\end{tabular}
\caption{
Model input parameters that can be set in \darknus. 
If a generic input parameter is set, the chosen model class is GenericHNLModel(), while if a three-portal model parameter is set, the chosen model class is ThreePortalModel(). 
It is not possible to mix the two.
All parameters are assumed to be real numbers, and satisfy $X_{ij} = X_{ji}$. Inputs are given by $X_{ij}$ with $i< j$. \label{tab:model_inputs}}
\end{table*}

\onecolumngrid

\footnotetext[1]{If specified while $g_\textup{D}$ is not, ignore default $g_\textup{D}$ and compute it from this parameter.}
\footnotetext[2]{If specified while $\varepsilon$ is not, ignore default $\varepsilon$ and compute it from this parameter.}
\footnotetext[3]{It is also possible to specify couplings to $\mathtt{f}=\mathtt{proton}$ and $\mathtt{f}=\mathtt{neutron}$, e.g. $\mathtt{cprotonV}$. In that case, the nucleon couplings override the calculation based on the sum of quark couplings such as in $\mathtt{cprotonV} = 2 \mathtt{cuV} + \mathtt{cdV}$.}

\twocolumngrid

\renewcommand{\arraystretch}{1.3}
\begin{table*}[t]
    \centering
    \begin{tabular}{|>{\tt}l|>{\em}c|>{\arraybackslash}p{11.5 cm}|}
    \hline
        \normalfont\textbf{Parameter} & \normalfont\textbf{Type} & \normalfont\textbf{Description} \\
    \hline
    \hline
        name & string & Name of the experiment (capital letters allowed); \\ 
        fluxfile & string & Path of the fluxes file with respect to either the experiment file directory or the program internal flux files directory; \\ 
        flux\_norm & float & Flux normalization factor: all fluxes should be normalized so that the units are $\mathrm{\#}\nu/(\mathrm{cm}^2 \, \mathrm{GeV} \, \mathrm{POT})$; \\ 
        erange & list of float & Neutrino energy range $E_\textup{min} \leq E \leq E_\textup{max}$ in $\mathrm{GeV}$; \\ 
        nuclear\_targets & string & Detector materials in the form of \texttt{"<element\_name><mass\_number>"}, e.g. \texttt{"Ar40"}; \\ 
        fiducial\_mass & float & Fiducial mass in $\mathrm{tons}$; \\ 
        fiducial\_mass\_per\_target & list of float & Fiducial mass for each target in the same order as specified in the \texttt{nuclear\_targets} parameter. \\ 
    \hline
    \end{tabular}
    \caption{The experiment parameters that can be set to define an experiment.
    These can only be set via the experimental definition files discussed in Sec.~\ref{sec:generator:input_exp}
    \label{tab:exp_parameters}}
\end{table*}
\makeatletter
\setlength{\@fptop}{2pt}
\makeatother
\renewcommand{\arraystretch}{1.3}
\begin{table*}[ht!]
    \centering
    \label{tab:pandas_format}
    \begin{tabular}{|>{\tt}l|>{\tt}l|c|>{\arraybackslash}p{11 cm}|}
    \hline
        \normalfont \textbf{Column} & \normalfont\textbf{Subcolumn} & \normalfont\textbf{Type} & \normalfont\textbf{Description} \\ \hline
    \hline
        {P\_projectile} & 0, 1, 2, 3 & \textit{float} & 4-momenta of beam neutrino \\ \hline
        {P\_decay\_N\_parent} & 0, 1, 2, 3 & \textit{float} & 4-momenta of HNL\_parent \\ \hline
        P\_target & 0, 1, 2, 3 & \textit{float} & 4-momenta of nucleus \\ \hline
        P\_recoil & 0, 1, 2, 3 & \textit{float} & 4-momenta of recoiled nucleus \\ \hline
        P\_decay\_ell\_minus & 0, 1, 2, 3 & \textit{float} & 4-momenta of e- \\ \hline
        P\_decay\_ell\_plus & 0, 1, 2, 3 & \textit{float} & 4-momenta of e+ \\ \hline
        P\_decay\_N\_daughter & 0, 1, 2, 3 & \textit{float} & 4-momenta of HNL\_daughter / nu\_daughter \\ \hline
        pos\_scatt & 0, 1, 2, 3 & \textit{float} & upscattering position \\ \hline
        pos\_decay & 0, 1, 2, 3 & \textit{float} & decay position of primary parent particle (N\_parent) -- secondary decay positions are not saved. \\ \hline
        w\_decay\_rate\_0 & ~ & \textit{float} & Weight of the decay rate of primary unstable particle: $\sum_i w_i = \Gamma_N$ \\ \hline
        w\_decay\_rate\_1 & ~ & \textit{float} & Weight of the decay rate of secondary unstable particle: $\sum_i w_i = \Gamma_X$ \\ \hline
        w\_event\_rate & ~ & \textit{float} & Weight for the event rate: $\sum_i w_i = $ event rate \\ \hline
        w\_flux\_avg\_xsec & ~ & \textit{float} & Weight of the flux averaged cross section: $\sum_i w_i =$ exposure$\times \int(\sigma \Phi_{\nu})$  \\ \hline
        target & ~ & \textit{string} & Target object, it will typically be a nucleus \\ \hline
        target\_pdgid & ~ & \textit{int} & PDG id of the target \\ \hline
        scattering\_regime & ~ & \textit{string} & Regime can be coherent, p-el, or n-el. \\ \hline
        helicity & ~ & \textit{string} & Helicity process: can be flipping or conserving; flipping is suppressed \\ \hline
        underlying\_process & ~ & \textit{string} & String of the underlying process, e.g, \texttt{"nu(mu) + proton\_in\_C12 -> N4 + proton\_in\_C12 -> nu(mu) + e+ + e- + proton\_in\_C12"} \\ \hline
    \end{tabular}
\caption{The format of the main \pandas dataframe used by \darknus. To access the negative lepton energy, for example, one can simply do \mintinline{python}|df["P_decay_ell_minus","0"]|.}
\end{table*}

\clearpage
\bibliographystyle{apsrev}
\bibliography{main}{}
\end{document}